\documentclass[11pt,a4paper,chicago]{amsart}
\pdfoutput=1
\usepackage[T1]{fontenc}
\usepackage[english]{babel}
\usepackage{fullpage}
\usepackage{graphicx}
\usepackage[numbers]{natbib}
\usepackage{amsmath, subfig}
\usepackage[amssymb]{SIunits}
\usepackage[colorlinks]{hyperref}

\newcommand{\ca}{Ca$^{2+}$}
\newcommand{\bc}{$\rm [Ca^{2+}]_i$}
\newcommand{\ipthree}{IP$_3$}
\newcommand{\ipr}{\ipthree R}

\newcommand{\iprOne}{type~I \ipr}
\newcommand{\iprTwo}{type~II \ipr}

\newcommand{\mthca}{\text{\ca}}
\newcommand{\mthipthree}{\text{\ipthree}}
\newcommand{\mthATP}{\text{ATP}}

\newcommand{\Jipr}{J_{\text{\ipr}}}
\newcommand{\Jryr}{J_{\text{RyR}}}
\newcommand{\Jserca}{J_{\text{SERCA}}}
\newcommand{\Jin}{J_{\text{in}}}
\newcommand{\Jpm}{J_{\text{pm}}}

\newcommand{\nO}{n_O}
\newcommand{\nC}{n_C}

\newcommand{\ullahstate}[3]{#1$_{#2}^{\text{#3}}$}

\newcommand{\pnull}{p_0}
\newcommand{\pO}{p_O}

\newcommand{\statDist}{\pi}

\DeclareMathOperator{\argmax}{argmax}

\usepackage{mathpazo}
\makeatletter
\newcommand{\addresseshere}{%
  \enddoc@text\let\enddoc@text\relax
}
\makeatother

\begin{document}

\title[Data-driven~\ipr\ modelling]{Data-driven modelling of the\\
  inositol trisphosphate receptor~(\ipr) and its role in\\
  calcium induced calcium release~(CICR)} 

\author[Ivo Siekmann, Pengxing Cao,James Sneyd \& Edmund~J.  Crampin]{Ivo Siekmann$^{1}$, Pengxing Cao$^2$, \\  James Sneyd$^3$ \& Edmund~J.  Crampin$^{1,2,4}$}

\address{
  \begin{minipage}[t]{1.0\linewidth}
    $^1$
    \begin{minipage}[t]{1.0\linewidth}
      \tiny Systems Biology Laboratory, Melbourne School of Engineering,\\
      University of Melbourne, Australia    
    \end{minipage}
\\
    $^2$
    \begin{minipage}[t]{1.0\linewidth}
      \tiny Department of Mathematics and Statistics,
      University of Melbourne, Australia    
    \end{minipage}
\\
    $^3$
    \begin{minipage}[t]{1.0\linewidth}
      \tiny Department of Mathematics,
      University of Auckland, New Zealand
    \end{minipage}
\\
    $^4$
    \begin{minipage}[t]{1.0\linewidth}
      \tiny School of Medicine,
      University of Melbourne, Australia    
    \end{minipage}
  \end{minipage}
}
\maketitle

\noindent\addresseshere 

% In particular we thought that your contribution does not need to be specific to the astrocyte but could rather present kinetic models for IP3R receptors (and/or RyRs) and for the relevant ones (such as IP3Rs of type 2 which are the predominant type in astrocytes) perhaps outlining possible techniques for model reduction in order to obtain simple reduced models for CICR signaling (then usable for astrocytes too).

\section{Introduction}
\label{sec:introduction}

A number of models have been published that relate different
physiological processes involving glial cells to calcium
dynamics. \citet{DeP:12a} give an overview of current problems in the
modelling of astrocytes. One area of continuing interest is the
propagation of signals between astrocytes via intercellular calcium
waves. \citet{Hoe:02a} investigated the spreading of signals between
astrocytes via calcium waves based on a model by
\citet{Sne:94a}. \citet{Ben:05a, Ben:06a} developed a more detailed
model of calcium waves that combines underlying calcium dynamics
with~ATP release by purinergic receptors in order to demonstrate that
calcium waves depend on ATP release rather than on \ipthree\ diffusion
through gap junctions as in the model by
\citet{Hoe:02a}. \citet{Edw:10a} later published a model that included
both modes of signal propagation and concluded that both were
necessary to account for data collected from the retina. Recently, the
study of calcium waves has been extended from one- or two-dimensional
to three-dimensional spatial domains \citep{Lal:14a}. \citet{Mac:13a}
model wave propagation on an astrocyte network derived from
experimental data. The Bennett et al. model was used for investigating spreading
depression, a wave of electrical silence that propagates through the
cortex and depolarises neurons and glial cells \citep{Ben:08a}. 

A fundamental problem in calcium dynamics in general is the question
how multiple signals can be encoded by the dynamics of a single
quantity, the concentration of calcium. \citet{DeP:08a,
  DeP:09a,DeP:09b} investigated how a stimulus could be encoded via
the frequency or the amplitude or both frequency and amplitude which
demonstrates that two different signals can be represented
independently in an individual calcium signal. \citet{Dup:11a} showed
in a detailed model how the signal received by a particular glutamate
receptor is encoded via calcium oscillations.

\citet{Lav:08a,Zen:09a, Rie:11a, Rie:11b} investigated spontaneous calcium
oscillations in astrocytes and \citet{Li:12a} explored their role in
spreading depression. 

Also the coupling of astrocyte network with the neural network has
been investigated. At the single-cell level, \citet{DeP:11a} modelled
the interaction of an astrocyte with a
synapse. \citet{All:09a,Pos:09a} study the influence of a network of
astrocytes on a neural network.

Most recently, \citet{Bar:14a, Bar:15a} explored the role of calcium
signalling in neural development.  By coupling calcium dynamics with a
model of the cell cycle they examine how glial progenitors
differentiate to neurons triggered by a calcium signal.

This review of the modelling literature on glial cells clearly
demonstrates that the importance of calcium dynamics is well
recognised---the majority of studies in the literature accounts for
calcium signalling and often models are used to find a link of
physiological processes with calcium signalling. In many cell types
including glial cells the inositol trisphosphate receptor (\ipr) plays
a crucial role in inducing oscillatory \ca\ signals. In the presence
of \ipthree, opening of \ipr\ channels leads to \ca\ release from the
endoplasmic reticulum (ER), an intracellular compartment with a very
high~\ca\ concentration a few orders of magnitude higher than that of
the cytoplasm. The~\ipr\ is activated by~\ca\ so that such a release
event dramatically increases the open probability of the~\ipr\ which
induces further release of \ca (henceforth called
calcium-induced-calcium-release, or CICR) until a high~\ca\
concentration in the channel environment eventually inhibits the~\ipr.

The Li-Rinzel model \citep{Li:94a}, an approximation of the classical
De Young-Keizer model \citep{DeY:92a}, is by far the most commonly
used representation of the~\ipr\ in models of glial cells. Only
\citet{All:09a} and \citet{Lav:08a} chose different models based on
\citet{Atr:93a} or \citet{Tu:05a}, respectively. \citet{Dup:11a} use
the model by \citet{Swi:94a} that explicitly accounts for the effect
of interactions in a cluster of~\ipr\ channels. Early models of
the~\ipr\ were designed to account for the bell-shaped~\ca\ dependency
of the open probability~$\pO$ of the channel described
by~\citet{Bez:91a}. Since then the dynamics of~\ipr\ in response to
varying concentrations of~\ipthree, \ca\ and ATP has been
characterised much more comprehensively as well as the differences
between the different isoforms of the~\ipr~(among the models mentioned
above, in fact, only \citet{Tu:05a} accounts for the fact that
astrocytes predominantly express type II \ipr).

The scope of current data-driven models of ion channels has advanced
beyond representing the average open probability~$\pO$. Recent models
capture the stochastic opening or closing of single~\ipr s\ in
aggregated Markov models i.e. instead of only modelling the stationary
behaviour of the channel they represent the dynamics of the~\ipr\
(Section~\ref{sec:ahmm}). Accurate representation of~\ipr\ dynamics
depends on various sources of experimental data
(Sections~\ref{sec:molecular}-\ref{sec:patch}) as well as appropriate
statistical methods for fitting Markov models to these
data~(Section~\ref{sec:estimation}). Statistical methods automate the
process of estimating parameters for a given Markov model. Thus, the
main challenge of data-driven ion channel modelling is to define the
structure of a Markov model which allows the integration of various
sources of experimental data. We illustrate this process with two
recent examples of models for the~\ipr~(Sections~\ref{sec:mechanistic}
and~\ref{sec:dataiprmodels}).

Once a model for a single channel has been developed, data from small
clusters of channels can be used to determine how well the behaviour
of a cluster is represented by an ensemble of single-channel models
(Section~\ref{sec:puffmodelling}). Studying the influence of an \ipr\
model on calcium dynamics allows us to evaluate the relative
importance of different aspects of single-channel
dynamics. \citet{Cao:14a} showed that the essential features of
calcium dynamics in airway smooth muscle could be preserved after
iteratively simplifying the \ipr\ model by \citet{Sie:12a} to a
two-state model that only accounted for the switching between the
inactive ``park'' and the active ``drive'' mode. In
Section~\ref{sec:IPRmodelreduction} it is shown that this also applies
to the puff distribution. This demonstrates that modal gating is the
most important regulatory mechanism of the \ipr. It also emphasises
that data-driven modelling of ion channels does not necessarily have
to lead to detailed models based on complicated model structures but
rather can be used so that relevant data is selected to represent ion
channels at the appropriate level of complexity for a given
application.

\section{Mathematical models of calcium dynamics/CICR}
\label{sec:camodels}

\begin{figure}
  \centering
\includegraphics[width=0.9\textwidth]{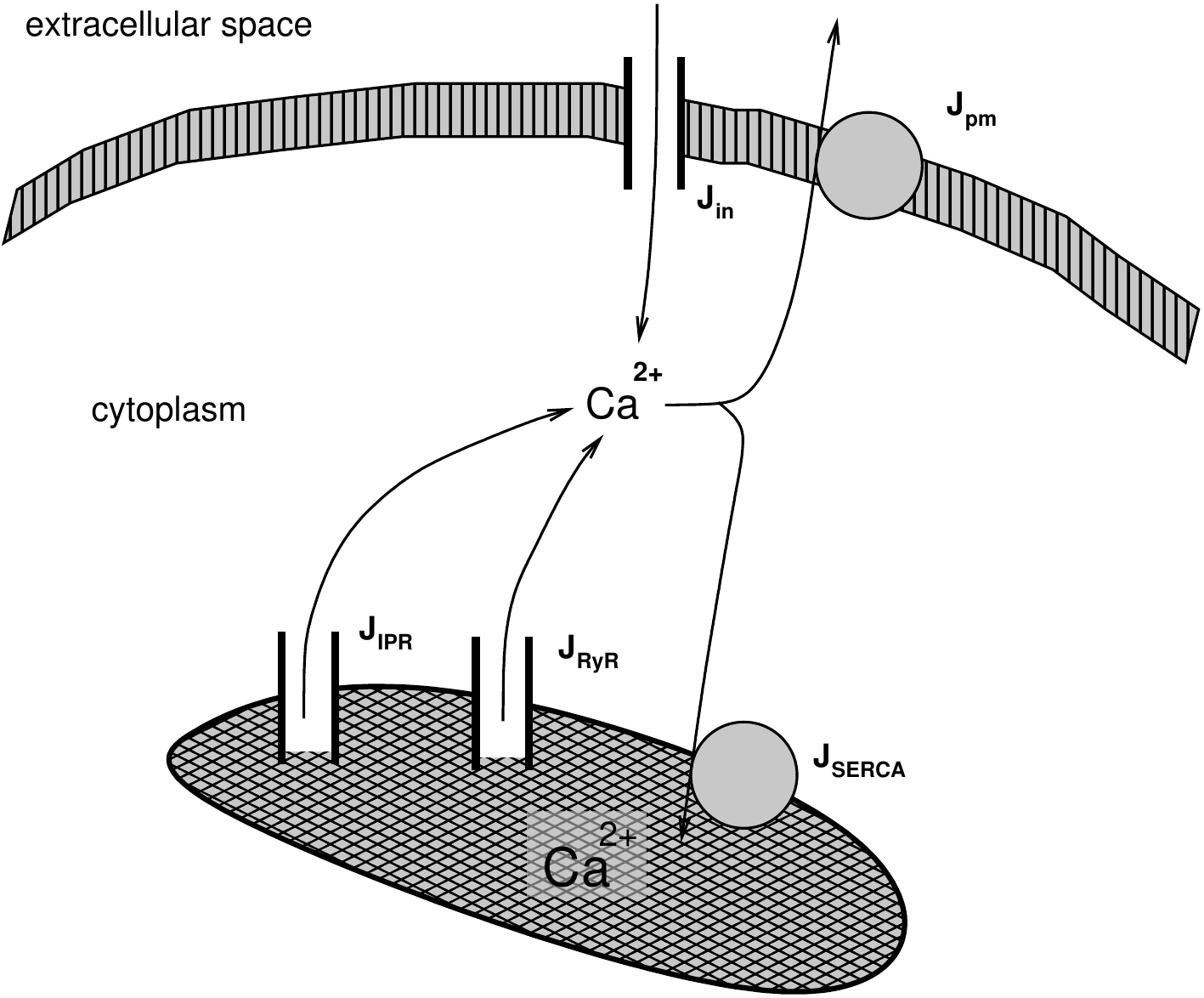}%
\caption{General structure of calcium fluxes in glial (and other
  non-excitable) cells. The central component is the
  flux~$J_{\text{IPR}}$ through the inositol trisphosphate
  receptor~(\ipr). The~\ipr\ is activated by binding of \ipthree\
  which is generated upon stimulation of the cell by an agonist. This
  causes the release of~\ca\ from the endoplasmic reticulum~(ER) to
  the cytoplasm. The resulting elevated~\ca\ concentration increases
  the open probability of the~\ipr\ and the ryanodine receptor~(RyR)
  which stimulates further~\ca\ release. This mechanism is known as
  calcium induced calcium release~(CICR). At high concentrations, \ca\
  inhibits the~\ipr, i.e. the open probability of the~\ipr\
  decreases. In consequence, $\Jserca$ influx into the ER through the
  SERCA pump dominates the efflux through~\ipr\ and RyR so that \ca\
  is reabsorbed by the ER. \ca\ exchange with the extracellular space
  is controlled by uptake through various channels~($\Jin$) and by
  extrusion via pumps~($\Jpm$).}
  \label{fig:cadynamics}
\end{figure}

The purpose of a mathematical model of CICR is to explain the
emergence of complex intracellular calcium dynamics such as
oscillations as the result of interdependent calcium fluxes. This
comprises both fluxes into and out of the cell as well as the exchange
between the cytosol and intracellular stores
(Figure~\ref{fig:cadynamics}).
The dynamics of cytosolic~($c$) and stored calcium~($c_\text{ER}$)
resulting from these fluxes can be represented by a system of
differential equations:

\begin{align}
  \label{eq:cicrC}
  \frac{d c}{d t} & = \Jipr +\Jryr + \Jin - \Jpm - \Jserca\\
  \label{eq:cicrCe}
  \frac{d c_\text{ER}}{d t} & = \gamma(\Jserca - \Jipr -\Jryr)
\end{align}

Here, $\Jin$ is calcium influx from the extracellular space via
calcium channels located in the cell membrane, $\Jpm$ accounts for
calcium removed from the cell by the plasma membrane pump. $\Jipr$
and~$\Jryr$ represent calcium release from the endoplasmic
reticulum~(ER) through the \ipr\ and the RyR, respectively,
and~$\Jserca$ stands for reuptake of calcium into the~ER by the SERCA
pump. The conversion factor~$\gamma$, the ratio of the cytoplasmic
volume to the ER volume, is necessary because calcium concentrations
are calculated with respect to the different volumes of these two
compartments. {The model~\eqref{eq:cicrC}, \eqref{eq:cicrCe} provides
  a description of~\ca\ concentrations across the whole cell. This
  means that we cannot account for spatial effects due to
  heterogeneities of the spatial distribution of~\ipr, SERCA and other
  relevant components of the system. By using a deterministic model we
  further assume that the various~\ca\ fluxes can be described as
  deterministic after averaging over a large number of channels and
  transporters. In Section~\ref{sec:using} we will consider a
  stochastic model over a small spatial domain for a cluster of
  interacting~\ipr s.}

In a whole-cell model of calcium dynamics such as~\eqref{eq:cicrC},
\eqref{eq:cicrCe}, a representation of the \ipr\ must, in principle,
just provide a functional expression for

\begin{equation}
  \label{eq:Jipr}
  \Jipr ([\mthipthree], [\mthca], [\mthATP]),
\end{equation}

the ligand-dependent flux through \ipr\ channels present in a
cell. {Because the calcium concentration~$[\mthca]$ is
  time-dependent, $\Jipr$ varies over time.} In the early days of
modelling of the \ipr, phenomenological models were used for
representing the \ipr\ flux. A good example is the model by
\citet{Atr:93a}:

\begin{equation}
  \label{eq:atri}
  \Jipr(p, c) = N_{\text{open}} k \left( \mu_0 + \frac{\mu_1}{k_\mu + p} \right)
  \left( b + \frac{V_1 c}{k_1 + c} \right) 
\end{equation}

where~$p=[\mthipthree]$, $c=[\mthca]$ and~$N_{\text{open}}$ is the
number of open channels. {The model by \citet{DeY:92a}
  is derived from more detailed assumptions on chemical interactions
  of the channel with its ligands. In Section~\ref{sec:mechanistic} we
  present a more recent model \citep{Ull:12a} that is representative
  for this approach.} The Hill function-type terms in~\eqref{eq:atri}
enabled Atri et al. to interpret their model in terms of a physical
process but the main motivation of the model was to obtain a fit of
the calcium-dependent whole-cell flux~$\Jipr$ to data collected by
\citet{Par:92a}. From a purely mathematical point of view,
phenomenological models seem to be the ideal approach for
investigating the role of \ipr\ in calcium dynamics---restriction to
minimal models that generate the desired behaviour ensures that model
behaviour can be analysed to a great extent. This allows us to test
hypotheses on \ipr\ regulation in an elegant way.

But the capability of simple mathematical expressions for the
macroscopic flux $\Jipr$ to perform the appropriate functional role in
calcium dynamics is only a relatively indirect test for \ipr\
models. By following a phenomenological approach we mostly ignore data
that gives more direct information on the \ipr, such as, for example,
the molecular structure of the channel protein which can be obtained
from crystallography and time series of opening and closing of a
single channel from patch-clamp recordings. Taking into account these
data may allow us to restrict the set of theoretically possible
mathematical expressions and, in this way, also the set of possible
mechanism.

\section{Data-driven modelling of single \ipr s}
\label{sec:data}

Because most biophysical data relate to single channels, data-driven
modelling involves an important conceptual step---instead of directly
specifying the whole-cell flux~$\Jipr$, we first construct a model for
the flux through a single channel. {Whereas for the macroscopic
  flux~$\Jipr$ which is averaged spatially over many channels
  distributed across the whole cell the deterministic
  model~\eqref{eq:Jipr} is appropriate, representing the flux through
  a single channel requires a stochastic model. In a second step,
  $\Jipr$ is then derived by appropriately averaging over the
  stochastic fluxes through individual channels.}

In Sections~\ref{sec:molecular} and~\ref{sec:patch} we describe two
sources of data that are commonly used for the construction of ion
channel models. \ca\ release data from small clusters of~\ipr,
so-called calcium puffs (Section~\ref{sec:puffs}), can be used for
validating models of single channels. In Section~\ref{sec:ahmm}
aggregated continuous-time Markov models, the mathematical framework
common to all models based on single-channel data, is introduced. A
short review of statistical approaches for fitting Markov models to
single-channel data is given in Section~\ref{sec:estimation}. In
Sections~\ref{sec:mechanistic} and~\ref{sec:dataiprmodels} examples of
two recent models of the~\ipr\ are given in order to illustrate
different modelling approaches. Earlier models have been reviewed by
\cite{Gin:09c} and \citet{Sne:05a}. Model comparisons \citep{Sne:04a,
  Hit:13a} generally show that models not parameterised by fitting to
experimental data may not do a very good job at reproducing the
statistical properties of ion channel kinetics.

\subsection{Molecular structure}
\label{sec:molecular}

The mathematical structure of many ion channel models is designed to
mimic the chemical structure of the channel protein. The motivation
for this approach is to link molecular structure of the ion channel to
its function. 

In vertebrates there exist three different genes encoding three
different types of the \ipr. In mammals, type~I~\ipr\ is ubiquitously
expressed but most cells express more than one isoform. The
predominant isoform in astrocytes is type~II~\ipr\ \citep{Sha:99a,
  Hol:02a}. For each isoform there are several splice variants.

Imaging the three-dimensional structure of the complete~\ipr\ and RyR
channel proteins is challenging and only recently have accurate 3D
visualisations of complete \ipr s using electron cryomicroscopy
(cryo-EM) become available \citep{Lud:13a}. Parts of the channel can
be imaged at higher resolution by crystallography and be superimposed
on cryo-EM images \citep{Fed:14a}. These studies have revealed that
\ipr\ channels are tetramers i.e. formed by binding of four \ipr\
proteins. These tetramers may consist of different \ipr\ subtypes but
experimental studies have so-far concentrated on investigating
homotetramers formed by four copies of the same subtype (but see
\citet{Alz:13a}). The classical model by \citet{DeY:92a} took into
account this information by building a model from identical subunits
that all had to be in an open state for the channel to open (although
the de Young-Keizer model assumed three instead of four subunits).

Analysis of the amino acid sequence by mutation experiments have
assigned functional roles to various segments, for example, the
\ipthree\ binding core~(IBC) which contains an \ipthree\ binding site
has been identified. There is less information on the number and
localisation of~\ca\ binding sites. Because localisation of \ca\
binding sites by mutation studies has been difficult, \citet{Fos:07a}
infer various \ca\ binding sensors from the observed co-regulation
by~\ipthree\ and \ca, see \citet{Fos:10a} for a summary. Often models
assume a certain number of~\ipthree\ and~\ca\ binding sites and
represent binding and unbinding of these ligands as transitions
between states regulated by mass action kinetics. This modelling
approach will be described in more detail in
Section~\ref{sec:mechanistic}.

\subsection{Patch-clamp recordings}
\label{sec:patch}

Detailed studies of individual ion channels became possible due to the
development of the patch-clamp technique. \citet{Neh:76a} were the
first to detect the flow of ions through a single ion channel by
measuring the resulting current at constant voltage. The time-course
of opening and closing can be inferred from the detected current which
stochastically jumps between zero (closed) and one or more small
non-zero current levels in the~pA range (open) whose sign
depends on the valence of the ion and the direction of the current.

\citet{Mak:15a} recently reviewed the single-channel literature
of~\ipr\ channels. An important experimental development that they
highlight relates to the difficulty that \ipr s are naturally
localised within cells rather than in the cell membrane. Whereas in
earlier patch-clamp experiments, \ipr\ channels were studied in
artificial lipid bilayers, more recently investigating \ipr\ in
isolated nuclei is favoured because it is assumed that nuclei provide
an environment similar to the endoplasmic reticulum~(ER), the native
domain of the~\ipr.

\subsubsection{Stationary data}
\label{sec:stationary}

If ligand concentrations (such as~\ipthree, \ca\ and~ATP) are kept
constant for the whole duration of the experiment we obtain stationary
data. These data allow us to observe the ``typical'' channel dynamics
for a given combination of ligands. The reason that we refer to these
data as ``stationary'' is that we assume that the channel has fully
adjusted to the concentration of ligands---the term stationary
suggests that the channel has reached its stationary probability
distribution, see Section~\ref{sec:ahmm}. Because the stationary
solution is only reached asymptotically we can, in theory, never be
sure that our ion channel has actually reached equilibrium. Instead we
can check if a data set is \emph{not} stationary by using indicators
such as the open probability. If the open probability averaged over a
sufficient number of data points spontaneously changes (which
indicates the switching of the channel to a different activity level)
the channel may exhibit modal gating.

\subsubsection{Modal gating}
\label{sec:modalgating}

Spontaneous switching between different levels of channel activity at
constant ligand concentrations has been observed for a long time. The
earliest example is perhaps from a classical study of the
large-conductance potassium channel~(BK) \citep{Mag:83a,Mag:83b}. In
\ipr\ channels modal gating was discovered only relatively recently
\citep{Ion:07a}. The authors found three different modes characterised
by high (H), intermediate (I) and low (L) levels of open
probabilities, $\pO^H$, $\pO^I$ and $\pO^L$. They also realised the
importance of modal gating for~\ipr\ regulation: they observed that
the same three modes seemed to exist for different combinations of
ligand concentrations. Because the \ipr\ mostly seemed to adjust the
time spent in each of the three modes they proposed that modal gating
is the major mechanism of ligand regulation in~\ipr\ channels.

One reason that the significance of modal gating has not been
appreciated until recently is due to the fact that switching between
different modes cannot always be recognised easily without statistical
analysis. Recently, \citet{Sie:14a} developed a statistical method
which for a given set of single-channel data detects switching between
an arbitrary number of modes~$M^i$ characterised by their respective
open probabilities~$\pO^{M^i}$. A software implementation which is
publicly available under
\url{https://github.com/merlinthemagician/icmcstat.git} was applied to
a large data set from \citet{Wag:12a}. \citet{Sie:14a} found that the
same two modes, an inactive ``park''~($\pO^{\text{park}} \approx 0\%$)
and an active ``drive'' mode~($\pO^{\text{drive}} \approx 70\%$), were
found across all combinations of ligands. There may be various reasons
why two modes were observed rather than the three modes found in the
earlier study \citep{Ion:07a}, see \citet{Sie:14a} for more
details. But more importantly, a detailed study of a bacterial
potassium channel (KscA) \citep{Cha:07a, Cha:07b, Cha:11a} strongly
suggests that the stochastic dynamics characteristic for each mode may
be closely related to distinct three-dimensional configurations
(conformations) of the channel. Thus, whereas it is often difficult to
relate individual open or closed states in ion channel models to
distinct conformations of the channel protein, the set of model states
that represents a particular mode may, in fact, have a biophysical
counterpart \citep{Sie:14a}. In order to confirm this hypothesis, more
studies of modal gating for a variety of channels are needed.

Independent from its biophysical significance, appropriately
accounting for modal gating is crucial from a modelling point of
view. As we will see in Section~\ref{sec:ahmm}, the phenomenon of
modal gating demonstrates that a Markov process must be observed for a
sufficiently long time in order to infer the correct stationary
distribution, otherwise we observe a ``quasi-steady state''.  For
example, a channel whose kinetics is restricted to an active and an
inactive mode can produce intermediate activity only by switching
between both modes. Thus, a model that is not capable of switching
between different levels of activity is misleading because it produces
a constant open probability instead of alternating between highly
different open probabilities. In their recent review \citet{Mak:15a}
explicitly recognise the importance of modal gating which so far has
only been represented in the most recent models \citep{Ull:12a,
  Sie:12a}.

\subsubsection{Response to rapid changes of ligand concentrations}
\label{sec:rapidperfusion}

Modal gating is an aspect of stationary data collected at constant
concentrations of ligands. In contrast, \citet{Mak:07a} designed an
experiment where~\ipthree\ and/or \ca\ concentrations in the channel
environment were rapidly altered in order to simulate an instantaneous
change of ligand concentrations. Switching from ligand concentration
where the~\ipr\ is inactive to conditions where the channel is
maximally activated (and vice versa), enabled \citet{Mak:07a} to
investigate the question how fast the~\ipr\ responds to varying ligand
concentrations. To illustrate the experiment let us consider the
change from inhibitory to activating conditions. At an inhibitory
condition, the open probability of the channel is very close to zero
$(\pO \approx 0)$ so that initially the~\ipr\ is most likely
closed. When changing from an inhibitory to an activating condition
the channel will activate but it needs a certain time to respond to
the change. In order to measure this latency, \citet{Mak:07a} recorded
the time the channel took from when they altered the ligand
concentration until the first opening. For the opposite change from
activating to inhibitory conditions they analogously detected the time
the channel needed to switch from a high to a low level of
activity. This experiment was repeated multiple times for switching
between the same conditions which enabled the authors to investigate
the latency statistics. It was not only discovered that for some
conditions the latencies were surprisingly long but interestingly,
they also found that for some conditions the latency distributions
were multi-modal which shows that multiple timescales may be observed
for the same latency.

Due to the substantial effort required to perform these experiments
(which have to be repeated multiple times for each condition where
each repeat only gives a single data point rather than a time course)
it is unsurprising that these data are very rare. In fact, to date,
\citet{Mak:07a} is the only data set of this kind currently
available. \citet{Mak:15a} explain that their data suggests that there
may be long refractory periods between release events from the same
\ipr\ which makes these results particularly relevant for the
modelling of \ca\ puffs. 

{
\subsection{Calcium puffs}
\label{sec:puffs}

So far we have only considered data recorded from single \ipr s. In
order to understand how the macroscopic flux~$\Jipr$ arises from the
release of many individual channels we have to consider the
hierarchical nature of \ca\ release. As reviewed by \citet{Par:96a,
  Fal:04a, Thu:12a} stochastic opening of a single \ipr\ channel leads
to a localised \ca\ release event (a \ca\ blip). Such a release
further sensitises neighbouring \ipr\ to induce more \ca\ release
through a few tightly clustered \ipr\ by CICR (a \ca\
puff). Sufficiently many puffs could eventually trigger a global
elevation of \bc\ that is able to propagate through the entire cell (a
\ca\ wave) \citep{Mar:99a}. Thus, \ca\ puffs play a crucial role that:
not only are puffs essential for the formation of functional global
\ca\ signals \citep{Cal:98a} but they also reflect the quantal \ca\
releases by stochastic openings of \ipr\ \emph{in vivo}
\citep{Smi:09a}.

Experimentally, \ca\ release at a specific spatial position can be
initiated by triggering release of caged \ipthree\ using a laser. A
relative measure for the local \ca\ concentration is obtained by
detecting fluorescent dye bound to~\ca\ using a light microscope. For
a given point within the cell the resulting time series is
characterised by a sequence of stochastic spikes that are highly
variable as far as the spike amplitude, the frequency and the time
interval between subsequent spikes, the inter-puff interval, is
concerned. From a modelling point of view, these data can be used to
test wether the single-channel behaviour represented in a model is
able to account for the release from a cluster of interacting \ipr
s. As explained in Section~\ref{sec:puffmodelling}, \citet{Cao:13a}
found that the original model by \citet{Sie:12a} was incapable of
generating the correct stochastic puff distribution as long as the
adaptation to different ligand concentrations was assumed to occur
instantaneously. After augmenting the model so that it accounted for
the latency data by \citet{Mak:07a} presented in the previous section
the puff statistics could be reproduced accurately.

}

The only other model that accounts for latency data is the model by
\citet{Ull:12a}. Because the models by \citet{Sie:12a,Cao:13a} and by
\citet{Ull:12a} are the only models that account for all aspects of
single-channel data assumed to be necessary for an understanding of
the~\ipr\ we focus on these two models and the alternative modelling
approaches that they represent in Sections~\ref{sec:mechanistic}
and~\ref{sec:dataiprmodels}.

\subsection{Aggregated continuous-time Markov models}
\label{sec:ahmm}

The most natural model for the stochastic process of opening and
closing of a single ion channel is the aggregated continuous-time
Markov model. A good introduction to the theory reviewed here is the
classical paper by \citet{Col:81a} which also gives some simple but
illustrative examples.

\begin{figure}[htbp]
  \centering
  \subfloat[\citet{Ull:12a}]{%
    \label{fig:Ullah}
    \includegraphics[height=0.45\textwidth]{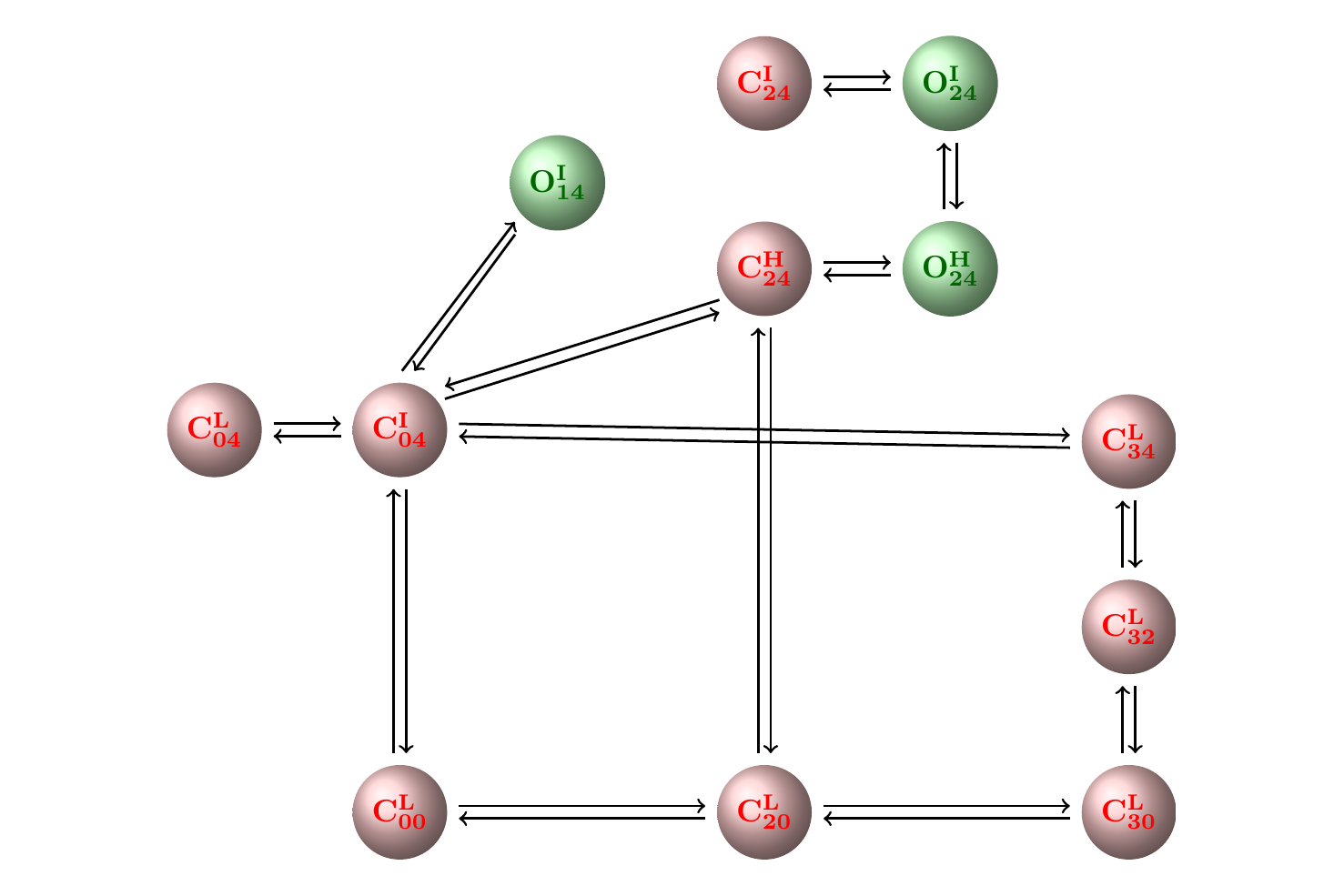}%
  }%
~  \subfloat[\citet{Sie:12a}]{%
  \label{fig:ParkDrive}
    \includegraphics[height=0.48\textwidth]{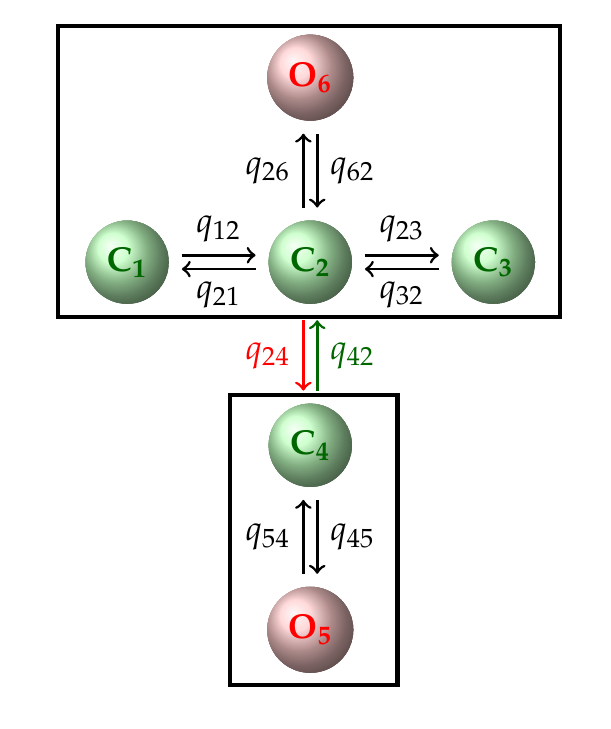}%
  }%
  \caption{Examples for recent Markov models of the~\ipr.}
  \label{fig:markov}
\end{figure}

An aggregated continuous-time Markov model is a graph on a set
of~$\nC$ closed and $\nO$ open
states~$S=\{C_1, \dots, C_{\nC}, O_{\nC+1}, \dots, O_{\nC+\nO}\}$
(Figure~\ref{fig:markov}).

Between adjacent states~$S_i$ and~$S_j$ the transition rate
(from~$S_i$ to~$S_j$) is given by~$q_{ij}>0$ so that the whole model
is represented by a matrix with constant coefficients, the
infinitesimal generator~$Q=(q_{ij})$. The time-dependent probability
distribution~$p(t)$ over the state set~$S$ is the solution of the
differential equation

\begin{equation}
  \label{eq:dQdt}
  \frac{d p(t)}{d t} = p(t) Q, \qquad p(0)=\pnull.
\end{equation}

The stochastic interpretation of~\eqref{eq:dQdt} is as follows: for a
given point in time, one particular state~$S_i$ of the model is
``active''.  But how long it will take until the current state~$S_i$
is vacated and which state~$S_j$ will be active after a time~$t$
cannot be answered with certainty (deterministically) due to the
stochastic transitions between states.

For the model defined by~\eqref{eq:dQdt} the Markov property holds
both for the stochastic sequence of active states as well as for the
time that it takes until the active state is left.

\begin{enumerate}
\item Which state~$S_j$ will be the next active state only depends on
  the currently active state~$S_i$, not on previously active states.
\item The time~$t_{S_i}$ it takes until the model exits from the
  state~$S_i$, also called the sojourn time in~$S_i$, does not depend
  on the time already spent in~$S_i$.
\end{enumerate}

The second point implies that sojourn times~$t_{S_i}$ must be
exponentially-distributed because the exponential distribution is the
only continuous probability distribution with this property. This
explains why multiple open and closed states may be needed for
accurately representing the opening and closing of ion channels. 

In order to ensure that~$p(t)$ is a stochastic vector
i.e. $\sum_{i=1}^{n_S} p_i, \; p_i \geq 0$ for
all~$t\geq 0$, the matrix~$Q$ must be conservative, i.e. for the
diagonal elements~$q_{ii}$ we have

\begin{equation}
  \label{eq:conservative}
  q_{ii}=- \sum_{j \neq i} q_{ij}, \qquad i,j=1, \dots, n_S. 
\end{equation}

Provided that~\eqref{eq:conservative} holds, the solution

\begin{equation}
  \label{eq:At}
  p(t)=\pnull \exp(Q t), 
\end{equation}

is a stochastic vector for all~$t>0$ if and only if the initial
distribution~$\pnull$ is a stochastic vector. From~\eqref{eq:At} the
time-dependent open probability $\pO(t)$ of the channel can be
calculated by summing over the individual probabilities of all open
states.

For large times~$t$ the solution~$p(t)$ approaches a stochastic
vector~$\statDist$ which is known as the stationary distribution. This
means that provided we wait sufficiently long, the expected frequency
of observing a state~$S_i$ approaches a
probability~$\statDist_i$. Because~$p(t)$ is the solution of a
differential equation, $\statDist$ is, in fact, a stationary solution
of~\eqref{eq:dQdt} i.e. can be obtained by solving the equation

\begin{equation}
  \label{eq:statdist}
  \statDist Q = 0.
\end{equation}

This homogeneous linear equation has non-trivial solutions because the
matrix~$Q$ is singular by~\eqref{eq:conservative}.  An argument based
on Perron-Frobenius theory for non-negative matrices ensures
that~$\statDist$ is a unique strictly positive stochastic
vector. Moreover, $\statDist$ is stable so that for~$t\to \infty$
indeed~$p(t)$ approaches~$\statDist$ i.e. we
have~$\lim_{t\to\infty} p(t) = \statDist$ \citep{Sen:81a}.

\subsection{Estimation of Markov models from experimental data}
\label{sec:estimation}

Whereas the mathematical framework of aggregated Markov models was
developed a short time after single channel data became available, the
statistical estimation of these models is a topic of current
research. Most commonly used are approaches based on Bayesian
statistics. For a given time series~$Y$ of open and closed events
recorded from an ion channel the conditional probability
density~$f(Q|Y)$, known as the posterior density in the Bayesian
framework, is used for determining a suitable Markov model with
infinitesimal generator~$Q$. Note that both~$Y$ and~$Q$ are considered
as random variables, thus the posterior distribution quantifies how
likely a model~$Q$ is under the condition that data~$Y$ have been
observed. Direct calculation of the posterior~$f(Q|Y)$ is analytically
intractable and computationally prohibitive but efficient approaches
for maximum likelihood estimation~(MLE) i.e. estimating

\begin{equation}
  \label{eq:mle}
  \hat{Q} = \argmax_{Q} f (Q|Y) 
\end{equation}

were published in the 1990s \citep{Qin:96a, Qin:97a,
  Col:96a}. Software implementations of these methods have been made
available freely for academic use. Currently, the methods by
\citet{Qin:96a, Qin:97a} can be obtained under the name \texttt{QUB}
as standalone GUI applications at
\url{http://www.qub.buffalo.edu/}. \texttt{DCPROGS} based on
\citet{Col:96a} is still under active development and the source code
of the most recent version has been published on github:
\url{https://github.com/DCPROGS}.

An alternative approach to maximum likelihood estimation has been
pursued since the late 1990s. The aim of Markov chain Monte
Carlo~(MCMC) is to approximate the posterior density~$f (Q|Y)$ by
sampling. MCMC enables us to randomly generate a
sequence~$(Q^k)_{k=1}^N$ of models such that the expected frequency of
a model~$Q^k$ within this sequence is as large as the
density~$f(Q^k|Y)$. Thus, by generating a sufficient number of
samples, the posterior~$f(Q|Y)$ is approximated.

The early method by \citet{Bal:99a} for estimation of a Markov
model~$Q$ depends on a suitable idealisation of discretely sampled
measurements to continuous open and closed times. This leads to a
difficult statistical problem that has been discussed widely in the
ion channel literature as the ``missed events'' problem. Rosales and
colleagues were the first to propose a method that directly uses the
discrete measurements and thus does not require further idealisation
of the data \citep{Ros:01a, Ros:04a}. Their algorithm estimates a
discrete-time Markov model which describes the transition
probabilities between states during a sampling interval rather than
the so-called infinitesimal generator~$Q$. \citet{Gin:09a} were the
first to propose a method for estimating~$Q$ from discretely-sampled
data, their method was extended to models with arbitrary numbers of
open and closed states by \citet{Sie:11a} and~\citet{Sie:12b}. The
current version of the software implementation of this method is
available on github:
\url{https://github.com/merlinthemagician/ahmm.git}. For an overview
of various approaches to statistical modelling based on single-channel
data, see \citet{Gin:09c}.

\begin{figure}
  \centering
  \subfloat{%
    \includegraphics[width=0.45\textwidth]{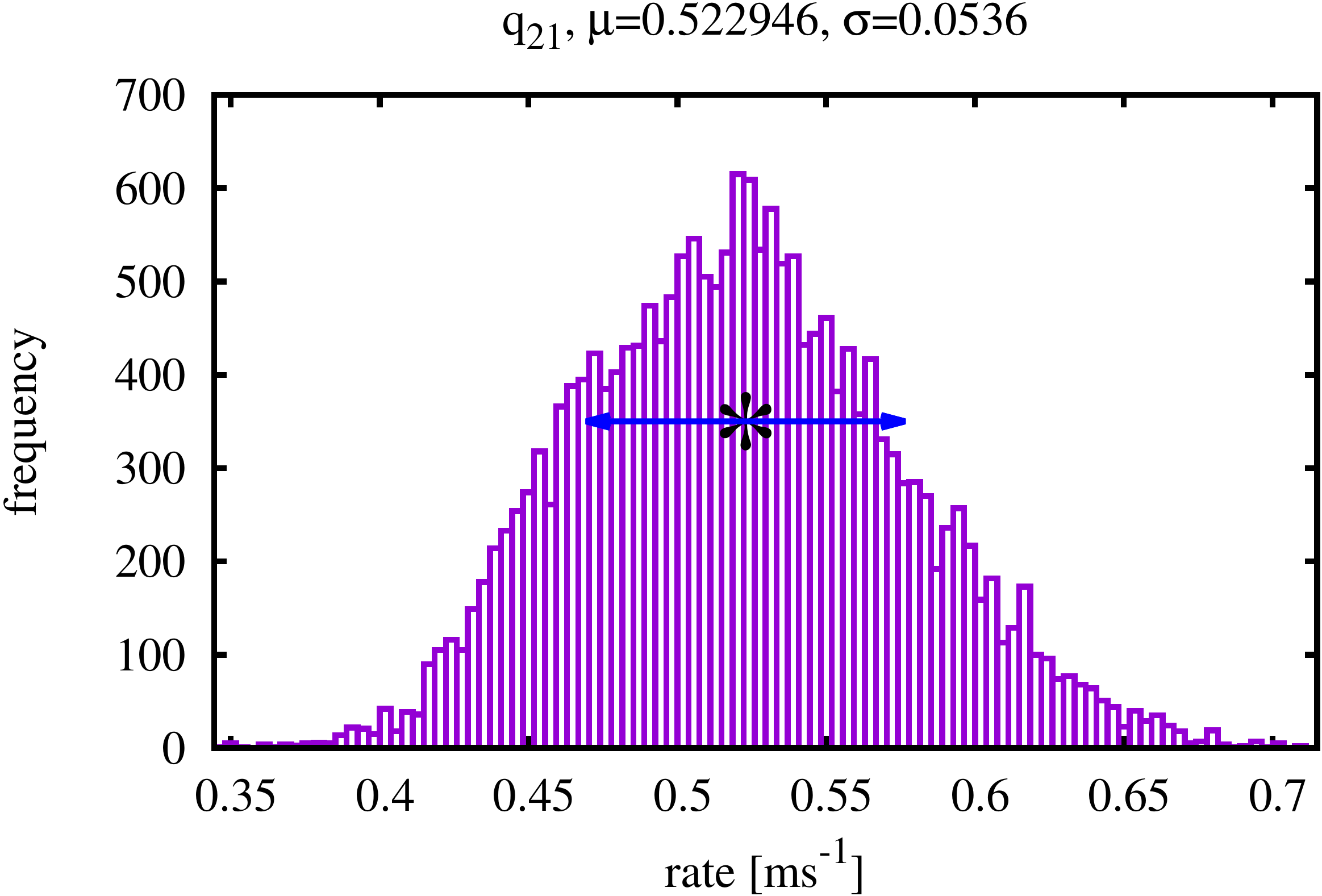}%
  }%
\;
\subfloat{%
  \includegraphics[width=0.45\textwidth]{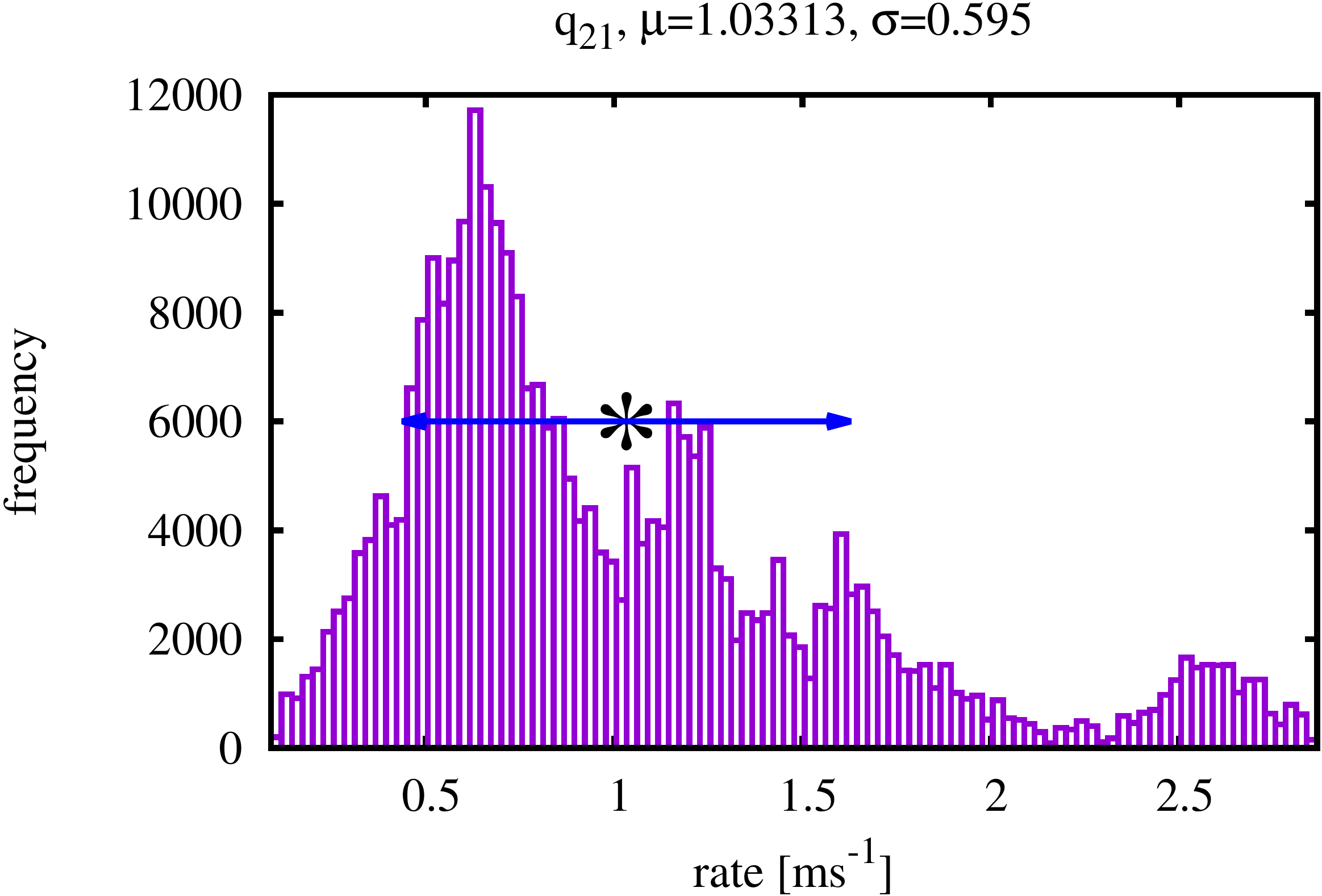}%
}%
\caption{Two examples for marginal distributions of rate
  constants. (a) shows a histogram with a well-defined mean~$\mu$ and
  a low standard deviation~$\sigma$ which indicates a low level of
  parameter uncertainty whereas the histogram in (b) shows a complex
  multi-modal distribution which shows that multiple values of the
  rate constants are capable of representing the data.}
  \label{fig:marginal}
\end{figure}

The crucial advantage of MCMC methods over MLE approaches is that
uncertainties can be comprehensively understood by analysing the
posterior~$f (Q|Y)$. Already marginal distributions
for individual rate constants~(Figure~\ref{fig:marginal}) are helpful
for localising and quantifying uncertainties within a model~$Q$.  %
But even more can be gained by analysing statistical relationships
between combinations of model parameters as, for example, demonstrated
by \citet{Sie:12b}. An important drawback of aggregated Markov models
is non-identifiability i.e. model structures whose parameters cannot
be inferred unambiguously from experimental data. Unfortunately,
non-identifiable aggregated Markov models have not been completely
classified \citep{Fre:85a,Fre:86a, Bru:05a}. But non-identifiability
can at least be detected by analysing the posterior
distribution~$f (Q|Y)$ \citep{Sie:12b}. Thus, MCMC allows us to
disentangle different causes of model uncertainty because it enables
us to distinguish between parameter uncertainties due to insufficient
or noisy data from pathologies in the structure of the model itself.

\subsection{The Ullah et al. model}
\label{sec:mechanistic}

A common approach for selecting a model structure for an ion channel
model (which goes back at least to the classical model by
\citet{DeY:92a}) is to identify the states of the Markov model with
different chemical states of the channel protein. As explained in
Section~\ref{sec:molecular}, the \ipr\ has various binding sites that
allow specific ligands such as \ca\ and \ipthree\ to bind to the
channel protein and induce conformational changes of its
three-dimensional structure. To account for this, model states are
distinguished by how many particles of each ligand are bound to the
channel. This assumption not only determines the state set of the
model but also the possible transitions between states---in each state
we can either bind a ligand to a free binding site or remove a ligand
from an occupied binding site. The dynamics of binding and unbinding
of ligands is modelled by the law of mass action so that, in
principle, the model is completely specified by the number of binding
sites for each ligand. However, in practice, such a model would be
heavily overparameterised when fitted to experimental data, so it is
necessary to simplify the model.

To illustrate this with an example, consider the recent model by
\citet{Ull:12a} which is representative for this approach. The model
states in Figure~\ref{fig:Ullah} are arranged in a grid whose
horizontal axis shows how many~\ca\ molecules are bound to the channel
and whose vertical axis indicates how many~\ipthree\ binding sites are
occupied. Thus, the position within the grid reflects for a specific
model state how many~\ca\ ions and how many \ipthree\ molecules,
respectively, are bound to the channel. For example, neither~\ca\
nor~\ipthree\ are bound to the state~\ullahstate{C}{00}{L} in the
lower left corner whereas two~\ca\ and four~\ipthree\ binding sites
are occupied for the states~\ullahstate{C}{{24}}{I},
\ullahstate{O}{24}{I}, \ullahstate{C}{24}{H} and
\ullahstate{O}{24}{H}. This is also indicated by the subscript
indices---the first digit stands for the number of~\ca\ ions whereas
the second digit accounts for the number of~\ipthree\ molecules bound
to the channel. Figure~\ref{fig:Ullah} shows that of the 20 possible
combinations of occupying~\ca, ATP and~\ipthree\ binding sites only a
subset of eight appears in the model. This considerable reduction is
due to the removal of ``low occupancy states''---\citet{Ull:12b}
developed a perturbation theory approach that allows them to omit
states with low stationary probabilities while at the same time
accounting for the delays caused by passing through these states.

The model is constructed in an iterative four step process integrating
several sources of data. In a first step, \citet{Ull:12a} use \ca\ and
\ipthree\ dependency of the average open probability~$\pO$ in order to
determine a minimal set of model states. By optimising an Akaike
information criterion (AIC) score function, a model with five closed,
\ullahstate{C}{00}{}, \ullahstate{C}{04}{}, \ullahstate{C}{24}{},
\ullahstate{C}{32}{} and \ullahstate{C}{34}{}, and one open state,
\ullahstate{O}{24}{}, was selected as the best fit for the~$\pO$ data.

In a second step, the ligand-dependent average probabilities~$\pi^L$,
$\pi^I$ and $\pi^H$ of being in modes characterised by three different
levels of activity as well as the open probabilities in each mode
($\pO^L$, $\pO^I$ and $\pO^H$) are used for assigning each of the six
model states with a mode. At this step, some additional states are
added because, for example, the state~\ullahstate{C}{04}{} must exist
both in the low (\ullahstate{C}{04}{L}) as well as the intermediate
mode (\ullahstate{C}{04}{L}) in order to get a good fit to the
data. To account appropriately for the~\ca\ dependency of~$\pO^I$, the
open probability in the intermediate mode, an additional
state~\ullahstate{O}{14}{I} had to be introduced.

In the first two steps, \citet{Ull:12a} use stationary probabilities
in order to determine which states should appear in the model without
considering transitions between states. In step 3 the authors infer
the transitions that are needed to account for the average sojourn
times~$\tau^L$, $\tau^I$ and~$\tau^H$ in the three modes whereas in
step 4, data on the \ipr\ response to rapid changes in~\ca\
and~\ipthree\ (latencies) is used for determining the remaining
transitions. Two additional states, \ullahstate{C}{20}{L} and
\ullahstate{C}{30}{L} are introduced in order to account for the
latency data.

Until this point, data is only used for determining the model
structure but not for parameter estimation. The model is finally
parameterised using the latency data from \citet{Mak:07a} or a
combination of these data and single-channel time series obtained at
three different constant \ca\ concentrations. 

\subsection{Siekmann et al. ``Park-Drive'' model}
\label{sec:dataiprmodels}

The main aims of the modelling study by \citet{Sie:12a} were first to
account for switching between an inactive ``park'' and an active
``drive'' mode observed in the data set by \citet{Wag:12a}. As
mentioned by \citet{Mak:07a} and \citet{Fos:10a, Mak:15a}, the
importance of modal gating is well-recognised and the implications for
not appropriately capturing the timescale separation of fast opening
and closing and slower switching between different activity levels is
obviously unsatisfactory from a modelling point of view.

Second, these data provided the possibility to build a model of two
different mammalian isoforms of the~\ipr, type~I and type~II~\ipr. In
addition to a comparative study of type~I and type~II~\ipr, these data
also include ligand-dependency of ATP in addition to \ipthree\ and
\ca.

Third, \citet{Sie:12a} followed a primarily statistical approach to
inference, rather than deriving the model from a binding scheme as the
model by \citet{Ull:12a} discussed above. Based on the experience of
the earlier study by \citet{Gin:09b} where similar data could be
fitted satisfactorily by a model with four states and only one
ligand-dependent pair of rate constants, the number of parameters
required to account for binding of \ipthree, \ca\ and ATP were likely
to lead to a highly overparameterised model.

Due to these considerations, \citet{Sie:12a} made the inactive
``park'' and the active ``drive'' mode the construction principle of
their model. In a first step, Markov models representing the
stochastic dynamics for these two modes were constructed based on
representative segments of the time series data that were
characteristic for one of the two modes. Models with different numbers
of states and model structures were fitted to these segments using the
method by \citet{Sie:11a, Sie:12b}. It was observed that the best fits
for either of the two modes across all combinations of ligands
available in the large data set by \citet{Wag:12a} were quantitatively
similar. This strongly suggested (consistent with \citet{Ion:07a})
that the dynamics within park and drive modes are ligand-independent
and that ligand-dependent regulation of~\ipr\ activity is achieved by
varying the prevalence of park or drive mode.

In a second step after both park and drive mode had been modelled
separately, a model of the ligand-dependent switching between the
ligand-independent modes was constructed. The structure for the full
Park-Drive model~(Figure~\ref{fig:ParkDrive}) was found by connecting
the Markov models of park and drive mode obtained previously with a
pair of transition rates. Due to the infrequent switching between park
and drive mode observed in the data it was decided that adding more
than a single pair of transition rates was statistically
unwarranted. The full Park-Drive model was then fitted to time series
for all combinations of ligands of the study by
\citet{Wag:12a}. The results of these fits established the
ligand-dependency of modal gating by the~\ipthree-, \ca- and
ATP-dependent variation of the two transition rates. 

Probably the most important result of this study is that only models
that take into account modal gating are able to accurately
capture~\ipr\ kinetics. A channel whose kinetics is restricted to an
active and an inactive mode can produce intermediate activity only by
switching between both modes. Thus, a model that is not capable of
switching between different levels of activity is misleading because
it produces a constant open probability instead of alternating between
highly different open probabilities. However, \citet{Cao:13a} showed
that accounting for modal gating alone was insufficient for modelling
stochastic~\ca\ release events (puffs) that arise from the
interactions of a few \ipr\ channels. This study showed that the
Park-Drive model has to be augmented by latency data \citep{Mak:07a}
in order to account for the delayed response of individual channels to
changes in ligand concentrations. 

Constructing the Park-Drive model based on the two modes proved very
useful in the study by \citet{Cao:14a}. The authors iteratively
reduced the Park-Drive model to a two-state model that only
approximates the dynamics of opening and closing {within} the modes
and focuses on the level of activity determined by the relative
prevalence of the modes. This further emphasises that switching
between park and drive mode rather than stochastic dynamics within the
modes is the most important mechanism of \ipr\ regulation. 

\subsection{Comparison of type~I and \iprTwo}
\label{sec:iprcomparison}

The experimental study by \citet{Wag:12a} not only investigated the
\ipr\ under a wide range of ligand conditions but also contrasted the
behaviour of type~I and \iprTwo. In the models for~type~I and \iprTwo\
constructed by \citet{Sie:12a} at a first glance the similarities
between both subtypes are probably more obvious than the
differences. First of all, it is striking that both \ipr\ subtypes can
not only be represented in the same model structure but that active
and inactive modes in both channels are nearly identical. This
indicates that both subtypes have the same modes and that their
differences are entirely due to differences in modal gating. 

One difference is that \iprTwo\ responds more sensitively to~\ipthree,
in contrast to \iprOne. The most important differences between both
subtypes was found to be~ATP regulation, see \citet{Wag:12a, Sie:12a}
for details.

\section{Using data-driven \ipr\ models in calcium dynamics}
\label{sec:using}

{So far we have focused on the dynamics of
  individual~\ipr s. In order to investigate the role of~\ipr s in
  calcium dynamics we will now consider the interaction of~\ipr s
  within a cluster.}

{
\subsection{Modeling calcium puffs using the Park-Drive \ipr\ model}
\label{sec:puffmodelling}

There is a large literature on stochastic models of calcium puffs for
which we refer to the recent review by \citet{Rue:14a}. Here we
present a simple model based on the Park-Drive model \citep{Sie:12a}
which is based on the following assumptions:
\begin{itemize}
\item The ER contains sufficiently high \bc\ to keep a nearly constant
  \ca\ release rate through a cluster of \ipr\ \citep{Ull:12c}. Thus,
  ER \bc\ dynamics is not explicitly modeled.
\item \ca\ fluxes through the cell membrane have little effect on the very localised \ca\ puffs far from cell membrane.
\item {We compartmentalise our model to capture
    heterogeneity within a cluster of \ipr s.  We assume that
    sufficiently far away from individual channels we have a
    homogeneous basal~\ca\ concentration~$c=$\bc\ that slowly responds
    to the total~\ca\ flux~$\Jipr$ through all~\ipr\ channels. In the vicinity
    of an open~\ipr\ channel this basal concentration~$c$ is elevated
    by a constant~$c_h$; once the channel closes it instantaneously
    equilibrates to the basal concentration~$c$.}
\end{itemize} 
Furthermore, \ca\ buffers are not considered except a \ca\ fluorescence dye. With these assumptions, the model is given as follows,
\begin{align}
  \label{eq:puffc}
  \frac{d c}{d t} & = \Jipr + J_{\rm leak}- \frac{V_dc}{c+K_d} - k_{\rm on}(B-b)c+k_{\rm off}b\\
  \label{eq:pufffluo}
  \frac{d b}{d t} & = k_{\rm on}(B-b)c - k_{\rm off}b
\end{align}
where $V_d c/(c + K_d)$ models the flux (mainly via diffusion and
SERCA) removing \ca\ from the puff site. $J_{\rm leak}$ represents
\ca\ leak current from the ER for stabilising the resting \bc\ of
$0.1 \mu M$ (a typical value). $B$ and $b$ represent the total dye
buffer concentration and \ca-bound dye buffer concentration
respectively, and the buffering process follows the mass action
kinetics. $\Jipr$ is the \ca\ flux through open \ipr, which is modeled
by the production of a constant release flux rate ($J_r$) and number
of open \ipr\ channels ($N_o$), i.e. $\Jipr = J_rN_o$. Each open \ipr\
will equally contribute to the elevation of cluster \bc, $c$. Note
that the actual \bc\ modulating each \ipr\ is either $c$ (when it is
in closed states) or $c+c_h$ (when it is in open states). Parameters
values are $J_r = 200\ \mu M/s$, $V_d = 4000\ \mu M/s$,
$K_d = 12\ \mu M$, $J_{\rm leak}= 33\ \mu M/s$, $B = 20\ \mu M$,
$k_{\rm on}= 150\ \mu M^{-1}s^{-1}$, $k_{\rm off}= 300\ s^{-1}$ and
$c_h = 120\ \mu M$ \citep{Cao:13a}. The cluster is assumed to contain
10 \ipr\ channels.

{The Park-Drive \ipr\ model is used to simulate \ipr\
  state and coupled to the deterministic equations via a
  hybrid-Gillespie method \citep{Rue:07a}.} However, the puff model based on the
Park-Drive model fails to reproduce nonexponential interpuff interval
(IPI) distribution due to the sole use of stationary single channel
data (i.e. \ca\ is fixed during measurement) in \ipr\ model
construction. This does not allow the model to capture the transient
single channel behaviour when \ca\ experiences a rapid change
\citep{Mak:07a,Cao:13a}. Thus, the Park-Drive model is modified by
incorporating time-dependent inter-mode transitions so that the
transient single channel behaviour is captured. In detail, the
transition rates $q_{24}$ and $q_{42}$ are changed from constants to
functions of four newly-introduced gating variables,
\begin{align}
  \label{eq:q24}
  q_{24} & = a_{24}+V_{24}(1-m_{24}h_{24})\\
  \label{eq:q42}
  q_{42} & = a_{42}+V_{42}m_{42}h_{42}
\end{align}
where $m_{24}$, $h_{24}$, $m_{42}$ and $h_{42}$ are gating variables obeying
\begin{equation}
\frac{dG}{dt} = \lambda_G(G_{\infty}-G),  \ G = m_{24}, h_{24}, m_{42}, h_{42}.
\end{equation}
$G_{\infty}$ is the steady state which is a function of channel-sensed
\ca\ and \ipthree\ concentrations and is determined by stationary
single channel data (i.e. the Park-Drive model). $\lambda_G$ is the
rate at which the steady state is approached. This is based on the
fact that a \ipr\ channel cannot immediately reach its steady state
upon a transient change in \ca\ concentration \citep{Mak:07a}. The
values of $\lambda_G$ for $m_{24}$, $h_{24}$ and $m_{42}$ are found to
be large so that the three gating variables could be approximated by
their steady states i.e. $G = G_{\infty}$, a method called
quasi-steady-state approximation.  However, $\lambda_{h_{42}}$ at low
\bc\ should be very small, reflecting a very slow recovery of \ipr\
from high \ca\ inhibition \citep{Mak:07a}. Note that when
$\lambda_{h_{42}}$ is sufficiently large, quasi-steady-state
approximation applies and the modified \ipr\ model reduces to the
original Park-Drive model. Details about the functions and parameters
can be seen in \citep{Cao:13a}.

\begin{figure}[htbp]
  \centering
  \includegraphics[scale = 0.31]{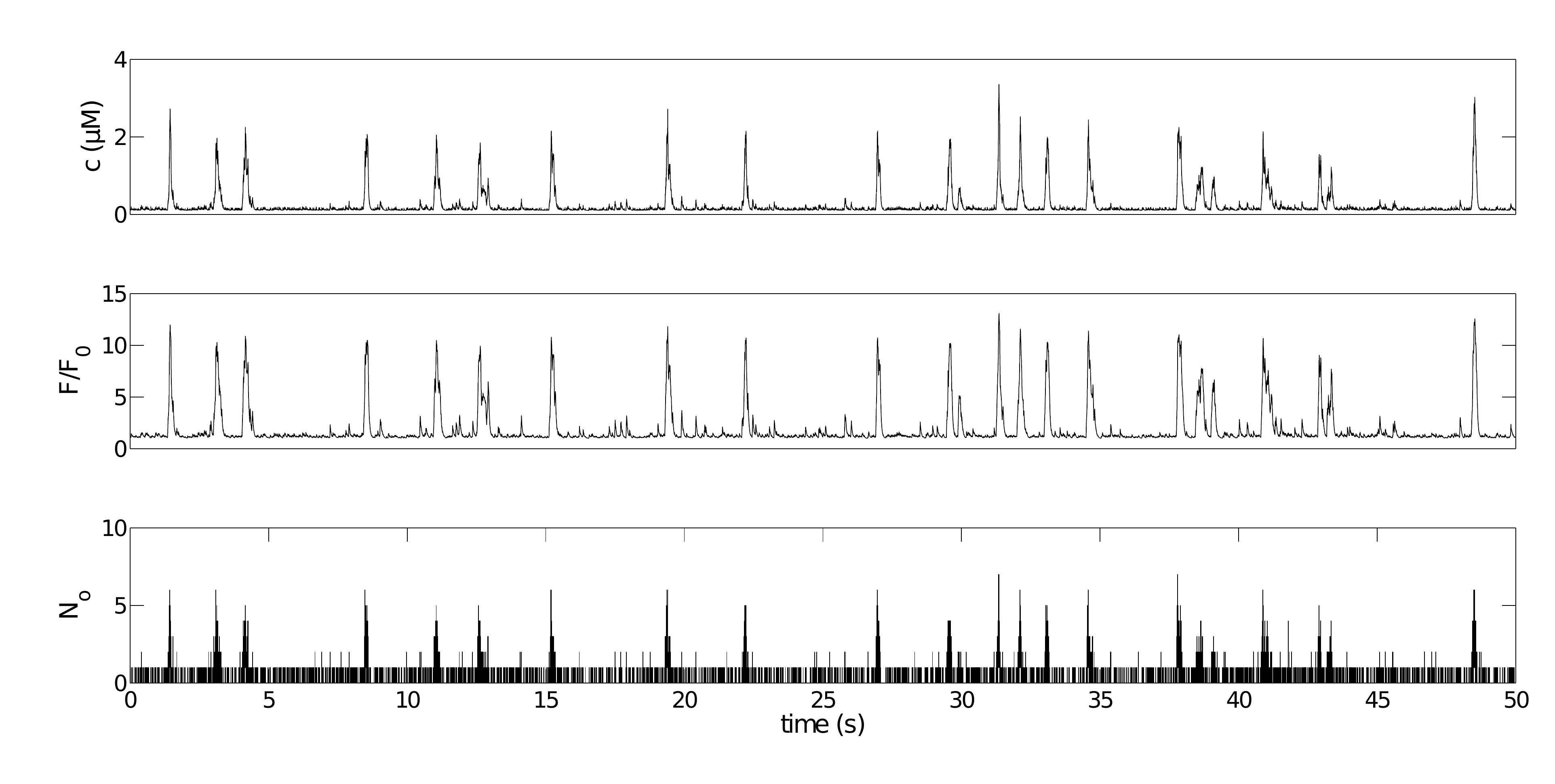}
  \caption{A simulation result of calcium puffs. $F/F_0$ represents the ratio of $b$ to its resting value. \ipthree\ concentration is $0.2 \mu M$. Adopted from \citep{Cao:13a}.}
  \label{fig:puffsim}
\end{figure}

An example of simulation results using the modified Park-Drive model is give in Fig.\ \ref{fig:puffsim}. The waiting time between two successive puffs (or interpuff interval, IPI) is a key statistics to quantify the underlying process governing the emergence of puffs. Fig.\ \ref{fig:IPIdis} shows that, as $\lambda_{h_{42}}$ at low \bc\ increases, the IPI distribution changes from nonexponential to exponential, demonstrating that the missing slow time scale in the original Park-Drive model is very crucial to explain the inhomogeneous Poisson process governing puff emergence found by (\citet{Thur:11a}). The IPI distributions were generated by fitting the probability density function proposed by \citet{Thur:11a} to the simulated IPI histograms \citep{Cao:13a}. The proposed IPI distribution is
\begin{equation}
  \label{eq:IPI}
  P = \lambda (1-e^{-\xi t})e^{[-\lambda t+\lambda(1-e^{-\xi t})/\xi]},
\end{equation}
where $t$ represent IPI. $\lambda$ is the puff rate, a measure of the typical IPI (similar to average puff frequency), and $\xi$ is the recovery rate.

\begin{figure}[htbp]
  \centering
  \includegraphics[scale = 0.55]{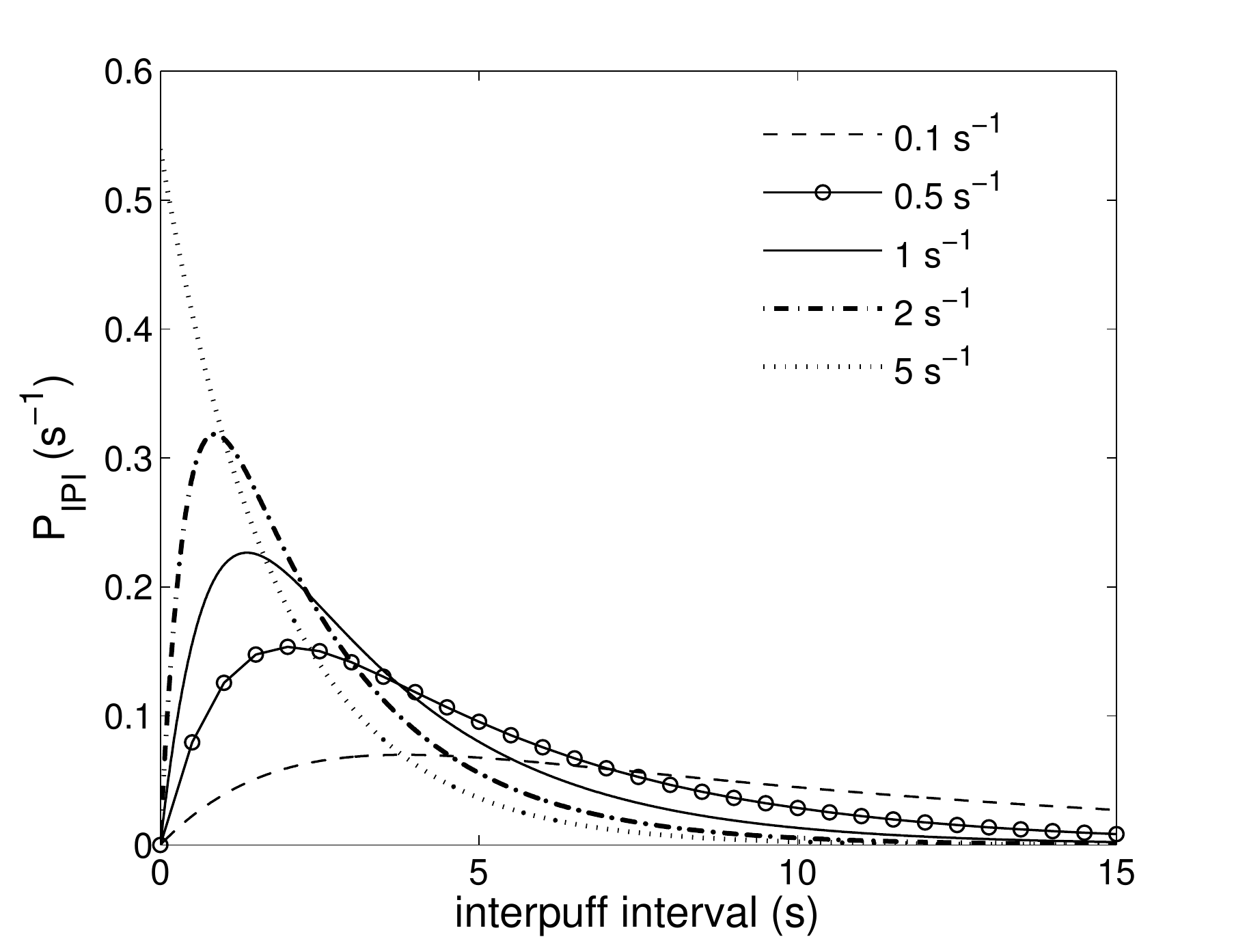}
  \caption{Dependence of IPI distribution on $\lambda_{h_{42}}$ (indicated in the legend) at low \bc. \ipthree\ concentration is $0.1 \mu M$. Adopted from \citep{Cao:13a}.}
  \label{fig:IPIdis}
\end{figure}

\begin{figure}[htbp]
  \centering
  \includegraphics[scale = 0.6]{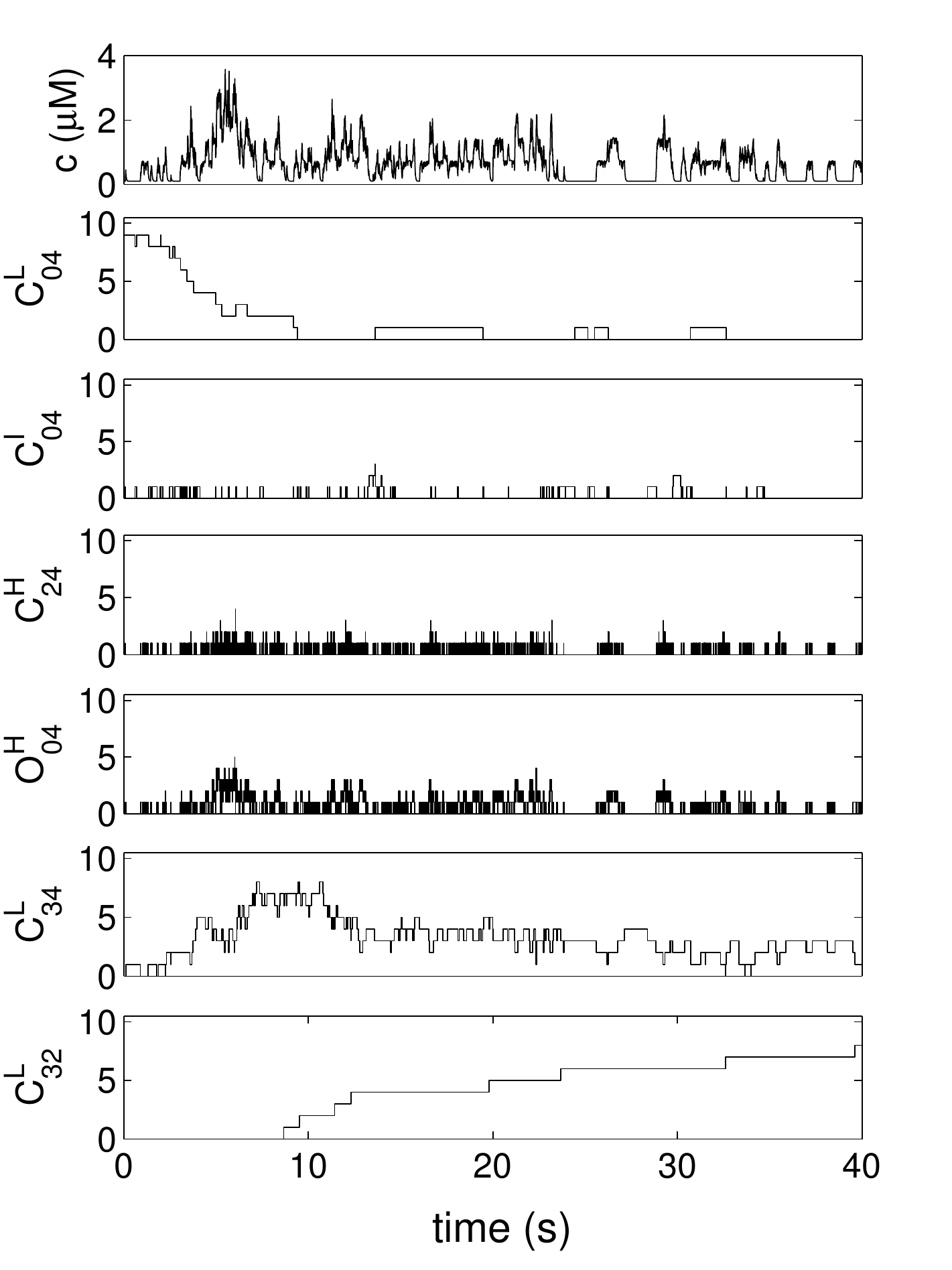}
  \caption{A simulation result of calcium puffs using the Ullah \ipr\ model. \ipthree\ concentration is $0.1 \mu M$. y-axis values indicate the number of \ipr\ channels in corresponding states. Parameter values for the puff model remain the same.}
  \label{fig:Ullahpuff}
\end{figure}

Hence, this example shows the particular importance of considering
both stationary and nonstationary data when constructing an \ipr\
model. However, even if a model is constructed based on both data
sets, it could also fail to reproduce \ca\ puffs. One example is the
Ullah model \citep{Ull:12a} as introduced in
Section~\ref{sec:mechanistic}. A model simulation using the same puff
model \eqref{eq:puffc}, \eqref{eq:pufffluo} with the Ullah model is
given in Fig.\ \ref{fig:Ullahpuff} where the \ca\ signal behaves very
irregularly and no puffs are clearly detected.
 
\subsection{The role of modal gating of \ipr\ in modulating calcium signals}
\label{sec:IPRmodelreduction}

The Park-Drive model (and its modified version) has the feature that
\ipr\ exist in two different modes, each of which contains multiple
states, some open, some closed. Intermode transitions are important
for modulating \ca\ signals because of their ligand- and
time-dependent property. However, structure within each mode may also
have substantial contribution to the formation of different \ca\
signals. Here, we examine the relative importance of intermode and
intramode transitions using model reduction methods. By reducing the
6-state \ipr\ model to a 2-state open/closed model, we will remove the
intramodal structure, and a direct comparison between the statistics
generated by the two \ipr\ models will show the importance of
intramodal structure.

The model reduction takes the following steps:
\begin{itemize}
\item The low probabilities of $C_1$ , $C_3$ and $O_5$ (sum of which is less than 0.03 for any \bc) means that the \ipr\ either rarely visit those states or have very short dwell time in those states. This allows to completely remove the three states from the 6-state model.

\item Transitions $q_{26}$ and $q_{62}$ are far larger (about two
  orders of magnitude) than $q_{24}$ and $q_{42}$. By taking a
  quasi-steady state approximation to the transition between $C_2$ and
  $O_6$, we have $O_6=C_2q_{26}/q_{62}$. Combining $C_2$ and $O_6$ to
  be a new state $D$, i.e. $D = C_2+O_6$, the 6-state model becomes a
  2-state model, where $D$ represents a partially open state with \ca\
  flux through the channel decreased by a factor of
  $q_{26}/(q_{62}+q_{26})$. Moreover, $q_{24}$ needs to be rescaled by
  $q_{62}/(q_{62}+q_{26})$ due to the quasi-steady state approximation
  so that the effective closing rate is
  $q_{24}q_{62}/(q_{62}+q_{26})$.
\end{itemize}

Fig.\ \ref{fig:6vs2} shows the distributions of interpuff interval, puff duration and amplitude generated by using the 6-state \ipr\ model (the Park-Drive model) and the reduced 2-state model. Reducing the intramodal structure does not qualitatively change the distributions but may lead to quantitative difference, which could be caused by missing open state $O_5$ that significantly contributes to the fluctuations of basal level of \bc. However, if the \ipr\ channel is not very sensitive to small fluctuations of basal \bc, the quantitative difference is significantly reduced \citep{Cao:14a}. Thus, the fundamental process governing the generation of \ca\ puffs and oscillations is primarily controlled by the modal structure but not the intramodal structure which improves the model fitting to the single-channel data.

\begin{figure}[htbp]
\centering
\subfloat[]{\includegraphics[scale=0.34]{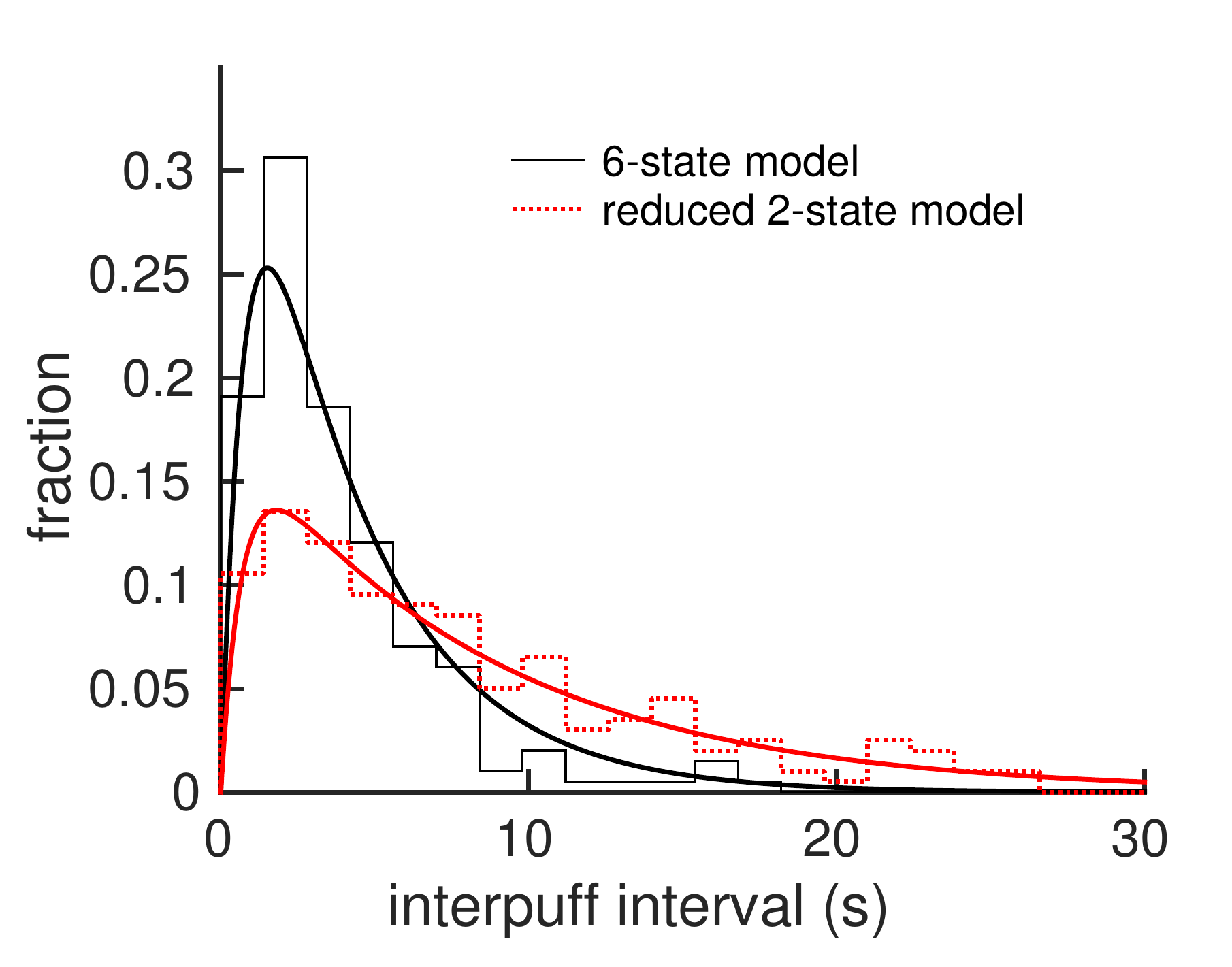}}
\subfloat[]{\includegraphics[scale=0.34]{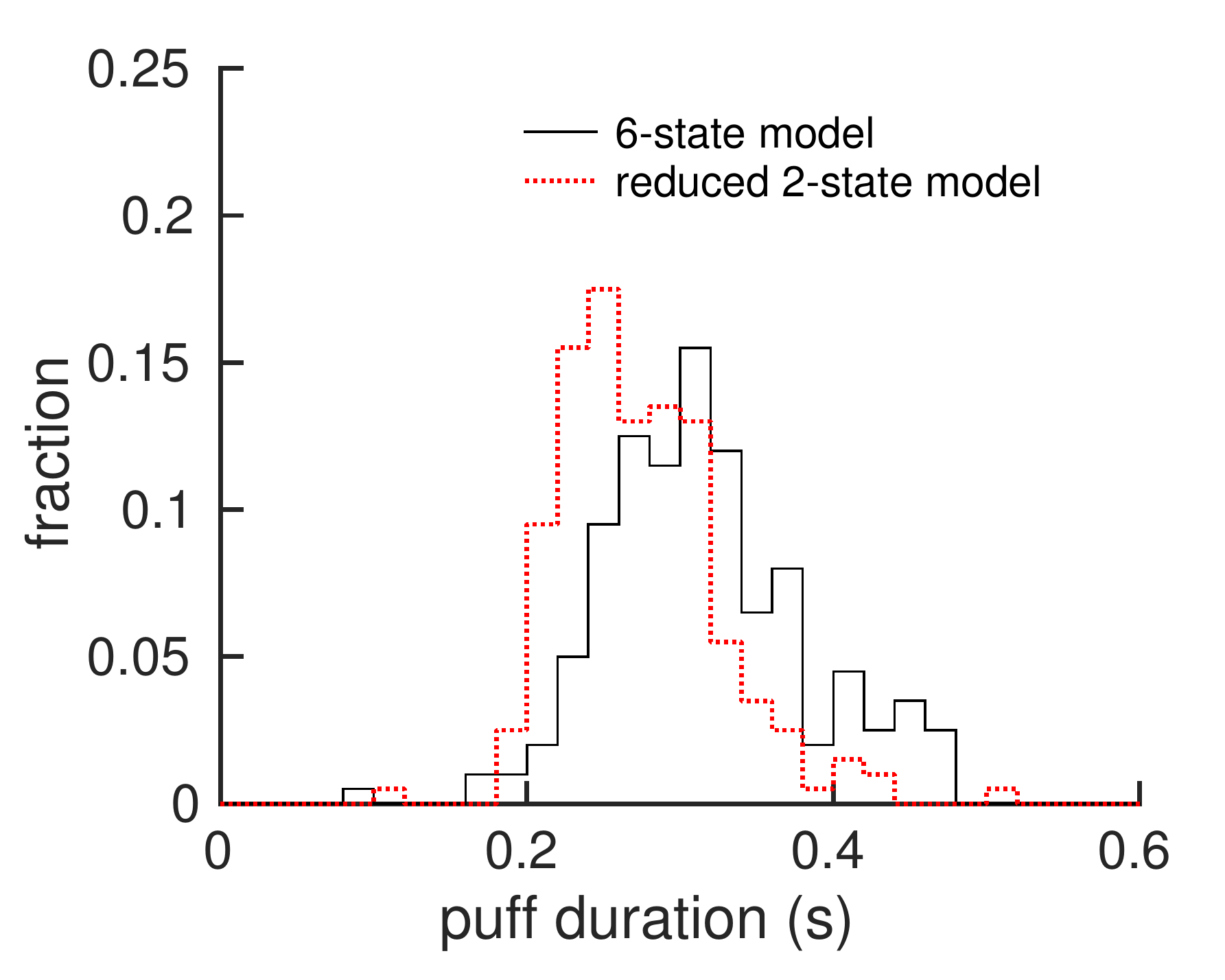}}\\
\subfloat[]{\includegraphics[scale=0.34]{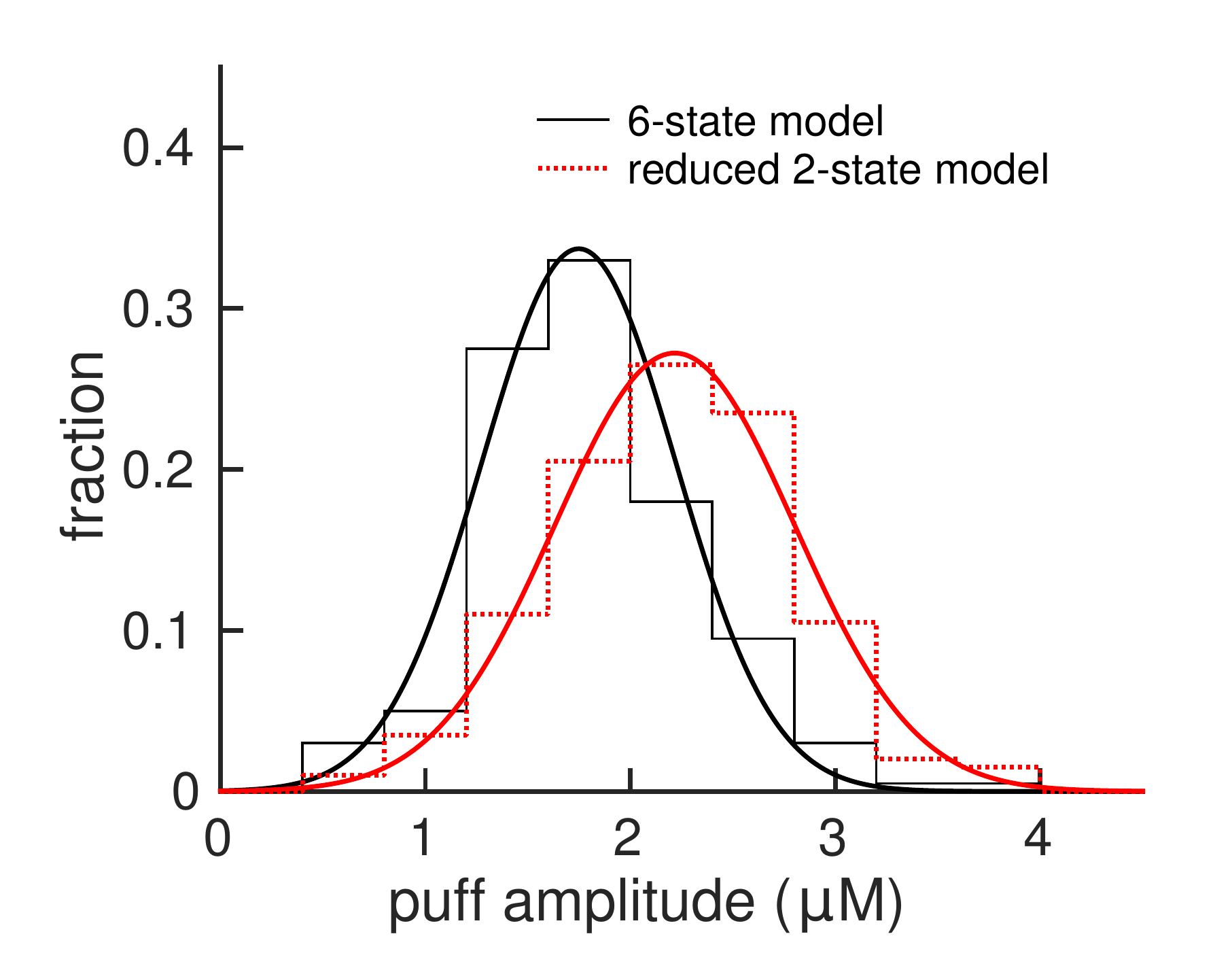}}
  \caption{Comparison of interpuff interval, puff duration and amplitude between the 6-state \ipr\ model (the Park-Drive model) and the reduced 2-state model. 199 samples for each model were used to generate {\bf A} and 200 samples for {\bf B} and {\bf C}. Interpuff interval distributions were fit by using Eq.\ref{eq:IPI} proposed by \citet{Thur:11a}. Puff amplitude distributions were fit by normal distribution.}
  \label{fig:6vs2}
\end{figure}

}

\section{Conclusions}
\label{sec:conclusions}

The \ipr\ plays a major role in CICR. For this reason, more and more
aspects of its behaviour have been investigated by experiments. It
usually turned out that new types of data had to be explicitly
included in a model to account for them. For example, in early models
such as the de Young-Keizer model \citep{DeY:92a}, the rate constants
were determined by fitting to the~$\pO$ observed at different calcium
concentrations. But it soon became obvious that models parameterised
with~$\pO$ data could not be used for extrapolating the channel
kinetics, i.e. the stochastic opening and closing. See \citet{Sne:05a}
or \citet{Ull:12a} for a more detailed explanation why it is
impossible to infer kinetics from the ligand dependency of the open
probability~$\pO$.

Just as kinetics cannot be inferred from~$\pO$ it turned out that the
response of the~\ipr\ to varying ligand concentrations cannot be
predicted from data collected at constant ligand concentrations. This
was demonstrated by the next generation of models that were directly
fitted to single-channel data, taking into account the stochastic
process of opening and closing.  The simplest assumption for
integrating models for different ligand concentration is that
the~\ipr\ adjusts instantaneously. If this was true we could represent
the channel kinetics appropriately by (to give a concrete example)
simply replacing the model for the kinetics at \unit{0.05}{\micro M}
with the model for the kinetics at \unit{0.2}{\micro M} calcium as we
increase the calcium concentration. But \citet{Cao:13a} showed that
only after taking into account rapid-perfusion data generated by
\citet{Mak:07a} was the model of \citet{Sie:12a} capable of generating
the correct puff distribution.

It is important to note that taking into account more data does not
necessarily have to lead to more complicated models. Instead, after
taking into account that the simpler kinetics of modal gating should
capture the part of the channel dynamics that is most important for
the functional role of the \ipr\ in CICR, \citet{Cao:14a} were able to
reduce the six-state model by \citet{Sie:12a} to a two-state
model. Thus, after interpreting experimental data in the right way, we
are able to build models for the functional role of \ipr\ that are
nearly as simple as the early phenomenological models.

\section{Future work}
\label{sec:future}

After reviewing the current state of data-driven approaches to
investigating the~\ipr\ we would like to take a look at promising
future directions. In order to address the particular importance of
modal gating, \citet{Sie:15a} develop a novel hierarchical model
structure that enables us to combine Markov models that represent the
stochastic switching between modes with models that account for the
characteristic opening and closing within different modes. Thus,
models for both processes can be fitted separately (e.g. using the
method by \citet{Sie:11a,Sie:12b}) after analysing the data with
statistical method presented by \citet{Sie:14a}.  This allows us to
build models for modal gating following a completely data-driven
approach.

More generally, we have compared two current models as representative
examples for different modelling approaches, the Ullah et
al. \citep{Ull:12a} and the Park-Drive model \citep{Sie:12a, Cao:13a,
  Cao:14a}. Although both approaches ultimately meet in the middle,
their different construction principles impose different requirements
for future progress. From a statistical point of view, representation
of ligand interactions with a channel by mass action kinetics as in
\citet{Ull:12a} defines a sufficiently large search space of
models. It is crucial to select from this search space an
appropriately simplified model that is obtained by removing states of
the full model in a consistent way. A method for model reduction is
provided by \citet{Ull:12b} and \citet{Ull:12a} demonstrate how data
can be used to statistically select from all possible simplified
models. A central principle of the biophysical approach is to design
models in a way that closely follows physical principles. In this
context, the bond-graph approach to modelling ion channels by
\citet{Gaw:14a, Gaw:15a} is highly relevant because it ensures that
physical principles are enforced when choosing a model structure.

For models that primarily focus on a statistically satisfying
representation in a first instance, the model selection problem arises
again but in the other direction. Rather than starting from a model
structure determined by an underlying mass action model,
\citet{Gin:09b} and \citet{Sie:12a} iteratively increased the number
of states in their model structure until further increasing the number
of parameters appears statistically unwarranted. This process is
time-consuming and may be computationally prohibitive if models exceed
a certain number of states. Developing a method that is able to
automatically compare models with an increasing number of states has
proven to be difficult indicated by the few number of studies that
have appeared on this subject after an early article on comparison of a
finite number of models \citep{Hod:99b}. A promising new direction is
the non-parametric Bayesian method developed by \citet{Hin:15a}
which allows the authors to estimate the number of states within an
ion-channel data set. Determining the required number of open and
closed states in a first step may increase efficiency because it
restricts the class of models which have to be compared in a second
step. 

\section*{Acknowledgements}
Funding from NIH grant R01-DE19245 is gratefully acknowledged. 

\bibliographystyle{chicago}
\bibliography{refer}

\begin{thebibliography}{}

\bibitem[\protect\citeauthoryear{Allegrini, Fronzoni, and Pirino}{Allegrini
  et~al.}{2009}]{All:09a}
Allegrini, P., L.~Fronzoni, and D.~Pirino (2009).
\newblock The influence of the astrocyte field on neuronal dynamics and
  synchronization.
\newblock {\em Journal of Biological Physics\/}~{\em 35\/}(4), 413--423.

\bibitem[\protect\citeauthoryear{Alzayady, Wagner, Chandrasekhar, Monteagudo,
  Godiska, Tall, Joseph, and Yule}{Alzayady et~al.}{2013}]{Alz:13a}
Alzayady, K.~J., L.~E. Wagner, R.~Chandrasekhar, A.~Monteagudo, R.~Godiska,
  G.~G. Tall, S.~K. Joseph, and D.~I. Yule (2013).
\newblock Functional inositol 1,4,5-trisphosphate receptors assembled from
  concatenated homo- and heteromeric subunits.
\newblock {\em Journal of Biological Chemistry\/}~{\em 288\/}(41),
  29772--29784.

\bibitem[\protect\citeauthoryear{Atri, Amundson, Clapham, and Sneyd}{Atri
  et~al.}{1993}]{Atr:93a}
Atri, A., J.~Amundson, D.~Clapham, and J.~Sneyd (1993).
\newblock A single-pool model for intracellular calcium oscillations and waves
  in the {Xenopus} laevis oocyte.
\newblock {\em Biophysical Journal\/}~{\em 65\/}(4), 1727--1739.

\bibitem[\protect\citeauthoryear{Ball, Cai, Kadane, and O'Hagan}{Ball
  et~al.}{1999}]{Bal:99a}
Ball, F.~G., Y.~Cai, J.~B. Kadane, and A.~O'Hagan (1999).
\newblock Bayesian inference for ion-channel gating mechanisms directly from
  single-channel recordings, using {M}arkov chain {M}onte {C}arlo.
\newblock {\em Proceedings of the Royal Society of London A\/}~{\em 455},
  2879--2932.

\bibitem[\protect\citeauthoryear{Barrack, Thul, and Owen}{Barrack
  et~al.}{2014}]{Bar:14a}
Barrack, D.~S., R.~Thul, and M.~R. Owen (2014).
\newblock Modelling the coupling between intracellular calcium release and the
  cell cycle during cortical brain development.
\newblock {\em Journal of Theoretical Biology\/}~{\em 347}, 17--32.

\bibitem[\protect\citeauthoryear{Barrack, Thul, and Owen}{Barrack
  et~al.}{2015}]{Bar:15a}
Barrack, D.~S., R.~Thul, and M.~R. Owen (2015).
\newblock Modelling cell cycle synchronisation in networks of coupled radial
  glial cells.
\newblock {\em Journal of Theoretical Biology\/}~{\em 377}, 85--97.

\bibitem[\protect\citeauthoryear{Bennett, Buljan, Farnell, and Gibson}{Bennett
  et~al.}{2006}]{Ben:06a}
Bennett, M.~R., V.~Buljan, L.~Farnell, and W.~G. Gibson (2006).
\newblock Purinergic junctional transmission and propagation of calcium waves
  in spinal cord astrocyte networks.
\newblock {\em Biophysical Journal\/}~{\em 91}, 3560--3571.

\bibitem[\protect\citeauthoryear{Bennett, Farnell, and Gibson}{Bennett
  et~al.}{2005}]{Ben:05a}
Bennett, M.~R., L.~Farnell, and W.~G. Gibson (2005).
\newblock A quantitative model of purinergic junctional transmission of calcium
  waves in astrocyte networks.
\newblock {\em Biophysical Journal\/}~{\em 89}, 2235--2250.

\bibitem[\protect\citeauthoryear{Bennett, Farnell, and Gibson}{Bennett
  et~al.}{2008}]{Ben:08a}
Bennett, M.~R., L.~Farnell, and W.~G. Gibson (2008).
\newblock A quantitative model of cortical spreading depression due to
  purinergic and gap-junction transmission in astrocyte networks.
\newblock {\em Biophysical Journal\/}~{\em 95}, 5648--5660.

\bibitem[\protect\citeauthoryear{Bezprozvanny, Watras, and
  Ehrlich}{Bezprozvanny et~al.}{1991}]{Bez:91a}
Bezprozvanny, I., J.~Watras, and B.~E. Ehrlich (1991).
\newblock Bell-shaped calcium-response curves of {Ins(1,4,5)P$_3$}-gated and
  calcium-gated channels from endoplasmic-reticulum of cerebellum.
\newblock {\em Nature\/}~{\em 351\/}(6329), 751--754.

\bibitem[\protect\citeauthoryear{Bruno, Yang, and Pearson}{Bruno
  et~al.}{2005}]{Bru:05a}
Bruno, W.~J., J.~Yang, and J.~E. Pearson (2005).
\newblock Using independent open-to-closed transitions to simplify aggregated
  {M}arkov models for ion channel gating kinetics.
\newblock {\em Proceedings of the National Academy of Science of the United
  States of America\/}~{\em 102\/}(16), 6326--6331.

\bibitem[\protect\citeauthoryear{Callamaras, Marchant, Sun, and
  Parker}{Callamaras et~al.}{1998}]{Cal:98a}
Callamaras, N., J.~S. Marchant, X.~P. Sun, and I.~Parker (1998).
\newblock Activation and coordination of {InsP$_3$}-mediated elementary
  {Ca$^{2+}$} events during global {Ca$^{2+}$} signals in {Xenopus} oocytes.
\newblock {\em Journal of Physiology\/}~{\em 509}, 81--91.

\bibitem[\protect\citeauthoryear{Cao, Donovan, Falcke, and Sneyd}{Cao
  et~al.}{2013}]{Cao:13a}
Cao, P., G.~Donovan, M.~Falcke, and J.~Sneyd (2013).
\newblock A stochastic model of calcium puffs based on single-channel data.
\newblock {\em Biophysical Journal\/}~{\em 105}, 1133--1142.

\bibitem[\protect\citeauthoryear{Cao, Tan, Donovan, Sanderson, and Sneyd}{Cao
  et~al.}{2014}]{Cao:14a}
Cao, P., X.~Tan, G.~Donovan, M.~J. Sanderson, and J.~Sneyd (2014, 08).
\newblock A deterministic model predicts the properties of stochastic calcium
  oscillations in airway smooth muscle cells.
\newblock {\em PLoS Computational Biology\/}~{\em 10\/}(8), e1003783.

\bibitem[\protect\citeauthoryear{Chakrapani, Cordero-Morales, Jogini, Pan,
  Cortes, Roux, and Perozo}{Chakrapani et~al.}{2011}]{Cha:11a}
Chakrapani, S., J.~F. Cordero-Morales, V.~Jogini, A.~C. Pan, D.~M. Cortes,
  B.~Roux, and E.~Perozo (2011).
\newblock On the structural basis of modal gating behaviour in {K$^+$}
  channels.
\newblock {\em Nature Structural and Molecular Biology\/}~{\em 18\/}(1),
  67--75.

\bibitem[\protect\citeauthoryear{Chakrapani, Cordero-Morales, and
  Peroso}{Chakrapani et~al.}{2007}]{Cha:07b}
Chakrapani, S., J.~F. Cordero-Morales, and E.~Peroso (2007).
\newblock A quantitative description of {KscA} gating~{II}: {Single-channel}
  currents.
\newblock {\em Journal of General Physiology\/}~{\em 130\/}(5), 479--496.

\bibitem[\protect\citeauthoryear{Chakrapani, Cordero-Morales, and
  Perozo}{Chakrapani et~al.}{2007}]{Cha:07a}
Chakrapani, S., J.~F. Cordero-Morales, and E.~Perozo (2007).
\newblock A quantitative description of {KscA} gating~{I}: {Macroscopic}
  currents.
\newblock {\em Journal of General Physiology\/}~{\em 130\/}(5), 465--478.

\bibitem[\protect\citeauthoryear{Colquhoun and Hawkes}{Colquhoun and
  Hawkes}{1981}]{Col:81a}
Colquhoun, D. and A.~G. Hawkes (1981).
\newblock On the stochastic properties of single ion channels.
\newblock {\em Proceedings of the Royal Society of London B\/}~{\em 211},
  205--235.

\bibitem[\protect\citeauthoryear{Colquhoun, Hawkes, and Srodzinski}{Colquhoun
  et~al.}{1996}]{Col:96a}
Colquhoun, D., A.~G. Hawkes, and K.~Srodzinski (1996).
\newblock Joint distributions of apparent open and shut times of single-ion
  channels and maximum likelihood fitting of mechanisms.
\newblock {\em Philosophical Transactions of the Royal Society of London
  A\/}~{\em 354}, 2555--2590.

\bibitem[\protect\citeauthoryear{De~Pitt\`a, Goldberg, Volman, Berry, and
  Ben-Jacob}{De~Pitt\`a et~al.}{2009}]{DeP:09a}
De~Pitt\`a, M., M.~Goldberg, V.~Volman, H.~Berry, and E.~Ben-Jacob (2009).
\newblock Glutamate regulation of calcium and {IP$_3$} oscillating and
  pulsating dynamics in astrocytes.
\newblock {\em Journal of Biological Physics\/}~{\em 35\/}(4), 383--411.

\bibitem[\protect\citeauthoryear{De~Pitt\`a, Volman, Berry, and
  Ben-Jacob}{De~Pitt\`a et~al.}{2011}]{DeP:11a}
De~Pitt\`a, M., V.~Volman, H.~Berry, and E.~Ben-Jacob (2011, 12).
\newblock A tale of two stories: Astrocyte regulation of synaptic depression
  and facilitation.
\newblock {\em PLoS Comput Biol\/}~{\em 7\/}(12), e1002293.

\bibitem[\protect\citeauthoryear{De~Pitt\`a, Volman, Berry, Parpura, Volterra,
  and Ben-Jacob}{De~Pitt\`a et~al.}{2012}]{DeP:12a}
De~Pitt\`a, M., V.~Volman, H.~Berry, V.~Parpura, A.~Volterra, and E.~Ben-Jacob
  (2012).
\newblock Computational quest for understanding the role of astrocyte signaling
  in synaptic transmission and plasticity.
\newblock {\em Frontiers in Computational Neuroscience\/}~{\em 6}.

\bibitem[\protect\citeauthoryear{De~Pitt\`a, Volman, Levine, and
  Ben-Jacob}{De~Pitt\`a et~al.}{2009}]{DeP:09b}
De~Pitt\`a, M., V.~Volman, H.~Levine, and E.~Ben-Jacob (2009).
\newblock Multimodal encoding in a simplified model of intracellular calcium
  signaling.
\newblock {\em Cognitive Processing\/}~{\em 10\/}(1), 55--70.

\bibitem[\protect\citeauthoryear{De~Pitt\`a, Volman, Levine, Pioggia, De~Rossi,
  and Ben-Jacob}{De~Pitt\`a et~al.}{2008}]{DeP:08a}
De~Pitt\`a, M., V.~Volman, H.~Levine, G.~Pioggia, D.~De~Rossi, and E.~Ben-Jacob
  (2008).
\newblock Coexistence of amplitude and frequency modulations in intracellular
  calcium dynamics.
\newblock {\em Physical Review E - Statistical, Nonlinear, and Soft Matter
  Physics\/}~{\em 77\/}(3).

\bibitem[\protect\citeauthoryear{De~Young and Keizer}{De~Young and
  Keizer}{1992}]{DeY:92a}
De~Young, G.~W. and J.~Keizer (1992).
\newblock A single-pool inositol 1,4,5-trisphosphate-receptor-based model for
  agonist-stimulated oscillations in {Ca$^{2+}$} concentration.
\newblock {\em Proceedings of the National Academy of Sciences\/}~{\em
  89\/}(20), 9895--9899.

\bibitem[\protect\citeauthoryear{Dupont, Lokenye, and Challiss}{Dupont
  et~al.}{2011}]{Dup:11a}
Dupont, G., E.~F.~L. Lokenye, and R.~J. Challiss (2011).
\newblock A model for {Ca$^{2+}$} oscillations stimulated by the type 5
  metabotropic glutamate receptor: {An} unusual mechanism based on repetitive,
  reversible phosphorylation of the receptor.
\newblock {\em Biochimie\/}~{\em 93\/}(12), 2132 -- 2138.

\bibitem[\protect\citeauthoryear{Edwards and Gibson}{Edwards and
  Gibson}{2010}]{Edw:10a}
Edwards, J. and W.~Gibson (2010).
\newblock A model for {Ca$^{2+}$} waves in networks of glial cells
  incorporating both intercellular and extracellular communication pathways.
\newblock {\em Journal of Theoretical Biology\/}~{\em 263\/}(1), 45--58.

\bibitem[\protect\citeauthoryear{Falcke}{Falcke}{2004}]{Fal:04a}
Falcke, M. (2004).
\newblock Reading the patterns in living cells -- the physics of {Ca$^{2+}$}
  signaling.
\newblock {\em Advances in Physics\/}~{\em 53\/}(3), 255--440.

\bibitem[\protect\citeauthoryear{Fedorenko, Popugaeva, Enomoto, Stathopulos,
  Ikura, and Bezprozvanny}{Fedorenko et~al.}{2014}]{Fed:14a}
Fedorenko, O.~A., E.~Popugaeva, M.~Enomoto, P.~B. Stathopulos, M.~Ikura, and
  I.~Bezprozvanny (2014).
\newblock Intracellular calcium channels: Inositol-1,4,5-trisphosphate
  receptors.
\newblock {\em European Journal of Pharmacology\/}~{\em 739}, 39 -- 48.
\newblock Special Issue on Calcium Channels.

\bibitem[\protect\citeauthoryear{Foskett and Mak}{Foskett and
  Mak}{2010}]{Fos:10a}
Foskett, J.~K. and D.-O.~D. Mak (2010).
\newblock Regulation of ip$_3$r channel gating by {Ca$^{2+}$} and {Ca$^{2+}$}
  binding proteins.
\newblock In I.~I. Serysheva (Ed.), {\em Structure and Function of Calcium
  Release Channels}, Volume~66 of {\em Current Topics in Membranes}, pp.\  235
  -- 272. Academic Press.

\bibitem[\protect\citeauthoryear{Foskett, White, Cheung, and Mak}{Foskett
  et~al.}{2007}]{Fos:07a}
Foskett, J.~K., C.~White, K.~Cheung, and D.~Mak (2007).
\newblock Inositol trisphosphate receptor {Ca$^{2+}$} release channels.
\newblock {\em Physiological Reviews\/}~{\em 87}, 593--568.

\bibitem[\protect\citeauthoryear{Fredkin, Montal, and Rice}{Fredkin
  et~al.}{1985}]{Fre:85a}
Fredkin, D.~R., M.~Montal, and J.~A. Rice (1985).
\newblock Identification of aggregated {M}arkovian models: {A}pplication to the
  nicotinic acetylcholine receptor.
\newblock In L.~M.~L. Cam and R.~A. Olshen (Eds.), {\em Proceedings of the
  Berkeley Conference in Honor of Jerzy Neyman and Jack Kiefer}, Volume~1,
  Belmont, CA, pp.\  269--289. Wadsworth.

\bibitem[\protect\citeauthoryear{Fredkin and Rice}{Fredkin and
  Rice}{1986}]{Fre:86a}
Fredkin, D.~R. and J.~A. Rice (1986).
\newblock On aggregated {M}arkov processes.
\newblock {\em Journal of Applied Probability\/}~{\em 23\/}(1), 208--214.

\bibitem[\protect\citeauthoryear{Gawthrop, Siekmann, and Crampin}{Gawthrop
  et~al.}{2015}]{Gaw:15a}
Gawthrop, P., I.~Siekmann, and E.~J. Crampin (2015).
\newblock A bond graph approach to chemoelectrical energy transduction in
  excitable membranes.
\newblock in preparation.

\bibitem[\protect\citeauthoryear{Gawthrop and Crampin}{Gawthrop and
  Crampin}{2014}]{Gaw:14a}
Gawthrop, P.~J. and E.~J. Crampin (2014).
\newblock Energy-based analysis of biochemical cycles using bond graphs.
\newblock {\em Proceedings of the Royal Society of London A: Mathematical,
  Physical and Engineering Sciences\/}~{\em 470\/}(2171).

\bibitem[\protect\citeauthoryear{Gin, Falcke, Wagner, Yule, and Sneyd}{Gin
  et~al.}{2009}]{Gin:09a}
Gin, E., M.~Falcke, L.~E. Wagner, D.~I. Yule, and J.~Sneyd (2009).
\newblock Markov chain {M}onte {C}arlo fitting of single-channel data from
  inositol trisphosphate receptors.
\newblock {\em Journal of Theoretical Biology\/}~{\em 257}, 460--474.

\bibitem[\protect\citeauthoryear{Gin, Falcke, {Wagner II}, Yule, and Sneyd}{Gin
  et~al.}{2009}]{Gin:09b}
Gin, E., M.~Falcke, L.~E. {Wagner II}, D.~I. Yule, and J.~Sneyd (2009).
\newblock A kinetic model of the inositol trisphosphate receptor based on
  single-channel data.
\newblock {\em Biophysical Journal\/}~{\em 96\/}(10), 4053--4062.

\bibitem[\protect\citeauthoryear{Gin, {Wagner II}, Yule, and Sneyd}{Gin
  et~al.}{2009}]{Gin:09c}
Gin, E., L.~E. {Wagner II}, D.~I. Yule, and J.~Sneyd (2009).
\newblock Inositol trisphosphate receptor and ion channel models based on
  single-channel data.
\newblock {\em Chaos: An Interdisciplinary Journal of Nonlinear Science\/}~{\em
  19\/}(3), 037104.

\bibitem[\protect\citeauthoryear{Hines, Bankston, and Aldrich}{Hines
  et~al.}{2015}]{Hin:15a}
Hines, K.~E., J.~R. Bankston, and R.~W. Aldrich (2015).
\newblock Analyzing single-molecule time series via nonparametric {Bayesian}
  inference.
\newblock {\em Biophysical Journal\/}~{\em 108\/}(3), 540 -- 556.

\bibitem[\protect\citeauthoryear{Hituri and Linne}{Hituri and
  Linne}{2013}]{Hit:13a}
Hituri, K. and M.-L. Linne (2013).
\newblock Comparison of models for {IP$_3$} receptor kinetics using stochastic
  simulations.
\newblock {\em PLoS ONE\/}~{\em 8\/}(4), e59618.

\bibitem[\protect\citeauthoryear{Hodgson and Green}{Hodgson and
  Green}{1999}]{Hod:99b}
Hodgson, M.~E.~A. and P.~J. Green (1999, {SEP 8}).
\newblock {Bayesian choice among Markov models of ion channels using Markov
  chain Monte Carlo}.
\newblock {\em Proceedings of the Royal Society of London Series A-Mathematical
  Physical and Engineering Sciences\/}~{\em {455}\/}({1989}), {3425--3448}.

\bibitem[\protect\citeauthoryear{H\"ofer, Venance, and Giaume}{H\"ofer
  et~al.}{2002}]{Hoe:02a}
H\"ofer, T., L.~Venance, and C.~Giaume (2002).
\newblock Control and plasticity of intercellular calcium waves in astrocytes:
  A modeling approach.
\newblock {\em The Journal of Neuroscience\/}~{\em 22\/}(12), 4850--4859.

\bibitem[\protect\citeauthoryear{Holtzclaw, Pandhit, Bare, Mignery, and
  Russell}{Holtzclaw et~al.}{2002}]{Hol:02a}
Holtzclaw, L., S.~Pandhit, D.~Bare, G.~Mignery, and J.~Russell (2002).
\newblock Astrocytes in adult rat brain express type 2 inositol
  1,4,5-trisphosphate receptors.
\newblock {\em GLIA\/}~{\em 39\/}(1), 69--84.

\bibitem[\protect\citeauthoryear{Ionescu, White, Cheung, Shuai, Parker,
  Pearson, Foskett, and Mak}{Ionescu et~al.}{2007}]{Ion:07a}
Ionescu, L., C.~White, K.-H. Cheung, J.~Shuai, I.~Parker, J.~E. Pearson, J.~K.
  Foskett, and D.-O.~D. Mak (2007).
\newblock Mode switching is the major mechanism of ligand regulation of
  {InsP$_3$} receptor calcium release channels.
\newblock {\em Journal of General Physiology\/}~{\em 130\/}(6), 631--645.

\bibitem[\protect\citeauthoryear{Lallouette, De~Pitt\`a, Ben-Jacob, and
  Berry}{Lallouette et~al.}{2014}]{Lal:14a}
Lallouette, J., M.~De~Pitt\`a, E.~Ben-Jacob, and H.~Berry (2014).
\newblock Sparse short-distance connections enhance calcium wave propagation in
  a {3D} model of astrocyte networks.
\newblock {\em Frontiers in Computational Neuroscience\/}~{\em 8}.

\bibitem[\protect\citeauthoryear{Lavrentovich and Hemkin}{Lavrentovich and
  Hemkin}{2008}]{Lav:08a}
Lavrentovich, M. and S.~Hemkin (2008).
\newblock A mathematical model of spontaneous {calcium(II)} oscillations in
  astrocytes.
\newblock {\em Journal of Theoretical Biology\/}~{\em 251\/}(4), 553--560.

\bibitem[\protect\citeauthoryear{Li, Chen, Zeng, Luo, and Li}{Li
  et~al.}{2012}]{Li:12a}
Li, B., S.~Chen, S.~Zeng, Q.~Luo, and P.~Li (2012).
\newblock Modeling the contributions of {Ca$^{2+}$} flows to spontaneous
  {Ca$^{2+}$} oscillations and cortical spreading depression-triggered
  {Ca$^{2+}$} waves in astrocyte networks.
\newblock {\em PLoS ONE\/}~{\em 7\/}(10).

\bibitem[\protect\citeauthoryear{Li and Rinzel}{Li and Rinzel}{1994}]{Li:94a}
Li, Y.-X. and J.~Rinzel (1994).
\newblock Equations for insp3 receptor-mediated {[Ca$^{2+}$]$_i$} oscillations
  derived from a detailed kinetic model: {A Hodgkin-Huxley} like formalism.
\newblock {\em Journal of Theoretical Biology\/}~{\em 166\/}(4), 461 -- 473.

\bibitem[\protect\citeauthoryear{Ludtke and Serysheva}{Ludtke and
  Serysheva}{2013}]{Lud:13a}
Ludtke, S.~J. and I.~I. Serysheva (2013).
\newblock Single-particle cryo-{EM} of calcium release channels: structural
  validation.
\newblock {\em Current Opinion in Structural Biology\/}~{\em 23}, 755--762.

\bibitem[\protect\citeauthoryear{Macdonald and Silva}{Macdonald and
  Silva}{2013}]{Mac:13a}
Macdonald, C. and G.~Silva (2013).
\newblock A positive feedback cell signaling nucleation model of astrocyte
  dynamics.
\newblock {\em Frontiers in Neuroengineering\/}~{\em 6}.

\bibitem[\protect\citeauthoryear{Magleby and Pallotta}{Magleby and
  Pallotta}{1983a}]{Mag:83b}
Magleby, K.~L. and B.~S. Pallotta ({1983}a).
\newblock {Burst kinetics of single calcium-activated potassium channels in
  cultured rat muscle}.
\newblock {\em {Journal of Physiology-London}\/}~{\em {344}}, {605--623}.

\bibitem[\protect\citeauthoryear{Magleby and Pallotta}{Magleby and
  Pallotta}{1983b}]{Mag:83a}
Magleby, K.~L. and B.~S. Pallotta ({1983}b).
\newblock {Calcium dependence of open and shut interval distributions from
  calcium-activated potassium channels in cultured rat muscle}.
\newblock {\em {Journal of Physiology-London}\/}~{\em {344}}, {585--604}.

\bibitem[\protect\citeauthoryear{Mak and Foskett}{Mak and
  Foskett}{2015}]{Mak:15a}
Mak, D.-O.~D. and J.~K. Foskett (2015).
\newblock Inositol 1,4,5-trisphosphate receptors in the endoplasmic reticulum:
  A single-channel point of view.
\newblock {\em Cell Calcium\/}~{\em 58\/}(1), 67 -- 78.

\bibitem[\protect\citeauthoryear{Mak, Pearson, Loong, Datta,
  Fern\'{a}ndez-Mongil, and Foskett}{Mak et~al.}{2007}]{Mak:07a}
Mak, D.-O.~D., J.~E. Pearson, K.~P.~C. Loong, S.~Datta,
  M.~Fern\'{a}ndez-Mongil, and J.~K. Foskett (2007).
\newblock Rapid ligand-regulated gating kinetics of single inositol
  1,4,5-trisphosphate receptor {Ca$^{2+}$} release channels.
\newblock {\em EMBO reports\/}~{\em 8\/}(11), 1044--1051.

\bibitem[\protect\citeauthoryear{Marchant, Callamaras, and Parker}{Marchant
  et~al.}{1999}]{Mar:99a}
Marchant, J., N.~Callamaras, and I.~Parker (1999).
\newblock Initiation of {IP$_3$}-mediated {Ca$^{2+}$} waves in {Xenopus}
  oocytes.
\newblock {\em EMBO Journal\/}~{\em 18}, 5285--5299.

\bibitem[\protect\citeauthoryear{Neher and Sakmann}{Neher and
  Sakmann}{1976}]{Neh:76a}
Neher, E. and B.~Sakmann (1976).
\newblock Single-channel currents recorded from membrane of denervated frog
  muscle fibres.
\newblock {\em Nature\/}~{\em 260\/}(5554), 799--802.

\bibitem[\protect\citeauthoryear{Parker, Choi, and Yao}{Parker
  et~al.}{1996}]{Par:96a}
Parker, I., J.~Choi, and Y.~Yao (1996).
\newblock Elementary events of {InsP3}-induced {Ca2+} liberation in xenopus
  oocytes: hot spots, puffs and blips.
\newblock {\em Cell Calcium\/}~{\em 20\/}(2), 105 -- 121.

\bibitem[\protect\citeauthoryear{Parys, Sernett, DeLisle, Snyder, Welsh, and
  Campbell}{Parys et~al.}{1992}]{Par:92a}
Parys, J.~B., S.~W. Sernett, S.~DeLisle, P.~M. Snyder, M.~J. Welsh, and K.~P.
  Campbell (1992).
\newblock Isolation, characterization, and localization of the inositol
  1,4,5-trisphosphate receptor protein in {Xenopus} laevis oocytes.
\newblock {\em The Journal of Biological Chemistry\/}~{\em 267\/}(26),
  18776--18782.

\bibitem[\protect\citeauthoryear{Postnov, Koreshkov, Brazhe, Brazhe, and
  Sosnovtseva}{Postnov et~al.}{2009}]{Pos:09a}
Postnov, D., R.~Koreshkov, N.~Brazhe, A.~Brazhe, and O.~Sosnovtseva (2009).
\newblock Dynamical patterns of calcium signaling in a functional model of
  neuron-astrocyte networks.
\newblock {\em Journal of Biological Physics\/}~{\em 35\/}(4), 425--445.

\bibitem[\protect\citeauthoryear{Qin, Auerbach, and Sachs}{Qin
  et~al.}{1996}]{Qin:96a}
Qin, F., A.~Auerbach, and F.~Sachs ({1996}).
\newblock {Idealization of single-channel currents using the segmental K-means
  method.}
\newblock {\em Biophysical Journal\/}~{\em {70}\/}({2, Part 2}), {MP432}.

\bibitem[\protect\citeauthoryear{Qin, Auerbach, and Sachs}{Qin
  et~al.}{1997}]{Qin:97a}
Qin, F., A.~Auerbach, and F.~Sachs ({1997}).
\newblock Maximum likelihood estimation of aggregated {M}arkov processes.
\newblock {\em Proceedings of the Royal Society of London Series B-Biological
  Sciences\/}~{\em {264}}, {375--383}.

\bibitem[\protect\citeauthoryear{Riera, Hatanaka, Ozaki, and Kawashima}{Riera
  et~al.}{2011}]{Rie:11a}
Riera, J., R.~Hatanaka, T.~Ozaki, and R.~Kawashima (2011).
\newblock Modeling the spontaneous {Ca$^{2+}$} oscillations in astrocytes:
  {Inconsistencies} and usefulness.
\newblock {\em Journal of Integrative Neuroscience\/}~{\em 10\/}(04), 439--473.

\bibitem[\protect\citeauthoryear{Riera, Hatanaka, Uchida, Ozaki, and
  Kawashima}{Riera et~al.}{2011}]{Rie:11b}
Riera, J., R.~Hatanaka, T.~Uchida, T.~Ozaki, and R.~Kawashima (2011).
\newblock Quantifying the uncertainty of spontaneous {Ca$^{2+}$} oscillations
  in astrocytes: {Particulars} of {Alzheimer's} disease.
\newblock {\em Biophysical Journal\/}~{\em 101\/}(3), 554--564.

\bibitem[\protect\citeauthoryear{Rosales}{Rosales}{2004}]{Ros:04a}
Rosales, R. (2004).
\newblock {MCMC} for {Hidden Markov Models} incorporating aggregation of states
  and filtering.
\newblock {\em Bulletin of Mathematical Biology\/}~{\em 66}, 1173--1199.

\bibitem[\protect\citeauthoryear{Rosales, Stark, Fitzgerald, and
  Hladky}{Rosales et~al.}{2001}]{Ros:01a}
Rosales, R., J.~A. Stark, W.~J. Fitzgerald, and S.~B. Hladky (2001).
\newblock {B}ayesian {R}estoration of {Ion Channel Records} using {Hidden
  Markov Models}.
\newblock {\em Biophysical Journal\/}~{\em 80\/}(3), 1088--1103.

\bibitem[\protect\citeauthoryear{R\"udiger}{R\"udiger}{2014}]{Rue:14a}
R\"udiger, S. (2014).
\newblock Stochastic models of intracellular calcium signals.
\newblock {\em Physics Reports\/}~{\em 534\/}(2), 39 -- 87.
\newblock Stochastic models of intracellular calcium signals.

\bibitem[\protect\citeauthoryear{R\"udiger, Shuai, Huisinga, Nagaiah, Warnecke,
  Parker, and Falcke}{R\"udiger et~al.}{2007}]{Rue:07a}
R\"udiger, S., J.~Shuai, W.~Huisinga, C.~Nagaiah, G.~Warnecke, I.~Parker, and
  M.~Falcke (2007).
\newblock Hybrid stochastic and deterministic simulations of calcium blips.
\newblock {\em Biophysical Journal\/}~{\em 93\/}(6), 1847 -- 1857.

\bibitem[\protect\citeauthoryear{Seneta}{Seneta}{1981}]{Sen:81a}
Seneta, E. (1981).
\newblock {\em Non-negative Matrices and Markov Chains\/} (2 ed.).
\newblock Springer Series in Statistics. New York: Springer.

\bibitem[\protect\citeauthoryear{Sharp, Nucifora~Jr., Blondel, Sheppard, Zhang,
  Snyder, Russell, Ryugo, and Ross}{Sharp et~al.}{1999}]{Sha:99a}
Sharp, A.~H., F.~C. Nucifora~Jr., O.~Blondel, C.~A. Sheppard, C.~Zhang, S.~H.
  Snyder, J.~T. Russell, D.~K. Ryugo, and C.~A. Ross (1999).
\newblock Differential cellular expression of isoforms of inositol
  1,4,5-triphosphate receptors in neurons and glia in brain.
\newblock {\em Journal of Comparative Neurology\/}~{\em 406\/}(2), 207--220.

\bibitem[\protect\citeauthoryear{Siekmann, Crampin, and Sneyd}{Siekmann
  et~al.}{2012}]{Sie:12b}
Siekmann, I., E.~J. Crampin, and J.~Sneyd (2012).
\newblock {MCMC} can detect non-identifiable models.
\newblock {\em Biophysical Journal\/}~{\em 103\/}(11), 1275--1286.

\bibitem[\protect\citeauthoryear{Siekmann, Fackrell, Taylor, and
  Crampin}{Siekmann et~al.}{2015}]{Sie:15a}
Siekmann, I., M.~Fackrell, P.~Taylor, and E.~J. Crampin (2015).
\newblock Modelling modal gating in ion channels with hierarchical {Markov}
  models.
\newblock in preparation.

\bibitem[\protect\citeauthoryear{Siekmann, Sneyd, and Crampin}{Siekmann
  et~al.}{2014}]{Sie:14a}
Siekmann, I., J.~Sneyd, and E.~J. Crampin (2014, June).
\newblock Statistical analysis of modal gating in ion channels.
\newblock {\em Proceedings of the Royal Society of London A\/}~{\em
  470\/}(2166), 20140030.

\bibitem[\protect\citeauthoryear{Siekmann, {Wagner II}, Yule, Crampin, and
  Sneyd}{Siekmann et~al.}{2012}]{Sie:12a}
Siekmann, I., L.~E. {Wagner II}, D.~Yule, E.~J. Crampin, and J.~Sneyd (2012).
\newblock A kinetic model of type {I} and type {II} {IP$_3$R} accounting for
  mode changes.
\newblock {\em Biophysical Journal\/}~{\em 103\/}(4), 658--668.

\bibitem[\protect\citeauthoryear{Siekmann, {Wagner II}, Yule, Fox, Bryant,
  Crampin, and Sneyd}{Siekmann et~al.}{2011}]{Sie:11a}
Siekmann, I., L.~E. {Wagner II}, D.~Yule, C.~Fox, D.~Bryant, E.~J. Crampin, and
  J.~Sneyd (2011).
\newblock {MCMC} estimation of {Markov} models for ion channels.
\newblock {\em Biophysical Journal\/}~{\em 100}, 1919--1929.

\bibitem[\protect\citeauthoryear{Smith and Parker}{Smith and
  Parker}{2009}]{Smi:09a}
Smith, I.~F. and I.~Parker (2009).
\newblock Imaging the quantal substructure of single {IP$_3$R} channel activity
  during {Ca$^{2+}$} puffs in intact mammalian cells.
\newblock {\em Proceedings of the National Academy of Sciences of the
  USA\/}~{\em 106\/}(15), 6404--6409.

\bibitem[\protect\citeauthoryear{Sneyd, Charles, and Sanderson}{Sneyd
  et~al.}{1994}]{Sne:94a}
Sneyd, J., A.~Charles, and M.~Sanderson (1994).
\newblock A model for the propagation of intercellular calcium waves.
\newblock {\em American Journal of Physiology - Cell Physiology\/}~{\em 266},
  C293--C302.

\bibitem[\protect\citeauthoryear{Sneyd and Falcke}{Sneyd and
  Falcke}{2005}]{Sne:05a}
Sneyd, J. and M.~Falcke (2005).
\newblock Models of the inositol trisphosphate receptor.
\newblock {\em Progress in Biophysics and Molecular Biology\/}~{\em 89},
  207--245.

\bibitem[\protect\citeauthoryear{Sneyd, Falcke, Dufour, and Fox}{Sneyd
  et~al.}{2004}]{Sne:04a}
Sneyd, J., M.~Falcke, J.~F. Dufour, and C.~Fox (2004).
\newblock A comparison of three models of the inositol trisphosphate receptor.
\newblock {\em Progress in Biophysics and Molecular Biology\/}~{\em 85},
  121--140.

\bibitem[\protect\citeauthoryear{Swillens, Combettes, and Champeil}{Swillens
  et~al.}{1994}]{Swi:94a}
Swillens, S., L.~Combettes, and P.~Champeil (1994).
\newblock Transient inositol 1,4,5-trisphosphate-induced {Ca2+} release: a
  model based on regulatory {Ca(2+)}-binding sites along the permeation
  pathway.
\newblock {\em Proceedings of the National Academy of Sciences\/}~{\em
  91\/}(21), 10074--10078.

\bibitem[\protect\citeauthoryear{Thurley, Skupin, Thul, and Falcke}{Thurley
  et~al.}{2012}]{Thu:12a}
Thurley, K., A.~Skupin, R.~Thul, and M.~Falcke (2012).
\newblock Fundamental properties of {Ca$^{2+}$} signals.
\newblock {\em Biochimica et Biophysica Acta\/}~{\em 1820\/}(8), 1185--1194.

\bibitem[\protect\citeauthoryear{{Thurley}, Smith, Tovey, Taylor, Parker, and
  Falcke}{{Thurley} et~al.}{2011}]{Thur:11a}
{Thurley}, K., I.~F. Smith, S.~C. Tovey, C.~W. Taylor, I.~Parker, and M.~Falcke
  (2011).
\newblock Timescales of {$\rm IP_3$-evoked $\rm Ca^{2+}$} spikes emerge from
  {$\rm Ca^{2+}$} puffs only at the cellular level.
\newblock {\em Biophysical Journal\/}~{\em 101}, 2638--2644.

\bibitem[\protect\citeauthoryear{Tu, Wang, and Bezprozvanny}{Tu
  et~al.}{2005}]{Tu:05a}
Tu, H., Z.~Wang, and I.~Bezprozvanny (2005).
\newblock Modulation of mammalian inositol 1,4,5-trisphosphate receptor
  isoforms by calcium: {A} role of calcium sensor region.
\newblock {\em Biophysical Journal\/}~{\em 88\/}(2), 1056--1069.

\bibitem[\protect\citeauthoryear{Ullah, Bruno, and Pearson}{Ullah
  et~al.}{2012}]{Ull:12b}
Ullah, G., W.~J. Bruno, and J.~E. Pearson (2012).
\newblock Simplification of reversible {Markov} chains by removal of states
  with low equilibrium occupancy.
\newblock {\em Journal of Theoretical Biology\/}~{\em 311}, 117--129.

\bibitem[\protect\citeauthoryear{Ullah, Mak, and Pearson}{Ullah
  et~al.}{2012}]{Ull:12a}
Ullah, G., D.-O.~D. Mak, and J.~E. Pearson (2012).
\newblock A data-driven model of a modal gated ion channel: {The} inositol
  1,4,5-trisphosphate receptor in insect {Sf9} cells.
\newblock {\em Journal of General Physiology\/}~{\em 140\/}(2), 159--173.

\bibitem[\protect\citeauthoryear{Ullah, Parker, Mak, and Pearson}{Ullah
  et~al.}{2012}]{Ull:12c}
Ullah, G., I.~Parker, D.~O.~D. Mak, and J.~E. Pearson (2012).
\newblock Multi-scale data-driven modeling and observation of calcium puffs.
\newblock {\em Cell Calcium\/}~{\em 52}, 152--160.

\bibitem[\protect\citeauthoryear{Wagner and Yule}{Wagner and
  Yule}{2012}]{Wag:12a}
Wagner, L.~E. and D.~I. Yule (2012).
\newblock Differential regulation of the {InsP$_3$} receptor type-1 and -2
  single channel properties by {InsP$_3$}, {Ca$^{2+}$} and {ATP}.
\newblock {\em The Journal of Physiology\/}~{\em 590\/}(14), 3245--3259.

\bibitem[\protect\citeauthoryear{Zeng, Li, Zeng, and Chen}{Zeng
  et~al.}{2009}]{Zen:09a}
Zeng, S., B.~Li, S.~Zeng, and S.~Chen (2009).
\newblock Simulation of spontaneous {Ca$^{2+}$} oscillations in astrocytes
  mediated by voltage-gated calcium channels.
\newblock {\em Biophysical Journal\/}~{\em 97\/}(9), 2429--2437.

\end{thebibliography}

\end{document}